\documentclass[12pt]{article}
\pdfoutput=1  
\usepackage{amsmath}
\usepackage{graphicx}
\usepackage{subfig}
\usepackage{slashed}

\def\LSP{\tilde{\chi}^0_1}

\def\chip{\tilde{\chi}_1^+}

\def\neu2{\tilde \chi_2^0}
\def\neu1{\tilde \chi_1^0}

\def\Re{{\cal R \mskip-4mu \lower.1ex \hbox{\it e}\,}}
\def\Im{{\cal I \mskip-5mu \lower.1ex \hbox{\it m}\,}}
\def\ie{{\it i.e.}}
\def\eg{{\it e.g.}}

\def\sub#1{_{\lower.25ex\hbox{$\scriptstyle#1$}}}
\def\tev{\,{\rm TeV}}
\def\gev{\,{\rm GeV}}

\def\dof{\,{\rm dof}}
\def\to{\rightarrow}

\def\subw{_{\rm w}}
\def\mh{\ifmmode m\sbl H \else $m\sbl H$\fi}
\def\mch{\ifmmode m_{H^\pm} \else $m_{H^\pm}$\fi}
\def\ztwo{\ifmmode Z_2\else $Z_2$\fi}
\def\zone{\ifmmode Z_1\else $Z_1$\fi}
\def\mtwo{\ifmmode M_2\else $M_2$\fi}
\def\mone{\ifmmode M_1\else $M_1$\fi}
\def\tb{\ifmmode \tan\beta \else $\tan\beta$\fi}
\def\xw{\ifmmode x\subw\else $x\subw$\fi}
\def\ch{\ifmmode H^\pm \else $H^\pm$\fi}
\def\lum{\ifmmode {\cal L}\else ${\cal L}$\fi}
\def\inpb{\ifmmode {\rm pb}^{-1}\else ${\rm pb}^{-1}$\fi}
\def\infb{\ifmmode {\rm fb}^{-1}\else ${\rm fb}^{-1}$\fi}
\def\epem{\ifmmode e^+e^-\else $e^+e^-$\fi}
\def\ppb{\ifmmode \bar pp\else $\bar pp$\fi}
\def\pbp{\ifmmode ~^(\bar p^)p\else $~^(\bar p^)p$\fi}
\def\bsg{\ifmmode B\to X_s\gamma\else $B\to X_s\gamma$\fi}
\def\bsll{\ifmmode B\to X_s\ell^+\ell^-\else $B\to X_s\ell^+\ell^-$\fi}
\def\bstt{\ifmmode B\to X_s\tau^+\tau^-\else $B\to
  X_s\tau^+\tau^-$\fi}

\newskip\zatskip \zatskip=0pt plus0pt minus0pt
\def\matth{\mathsurround=0pt}

\def\gsim{\mathrel{\mathpalette\atversim>}}
\def\atversim#1#2{\lower0.7ex\vbox{\baselineskip\zatskip\lineskip\zatskip
  \lineskiplimit
  0pt\ialign{$\matth#1\hfil##\hfil$\crcr#2\crcr\sim\crcr}}}

\def\sigv{\ifmmode \langle\sigma v\rangle\else $\langle\sigma
  v\rangle$\fi}
\def\tsigv{\ifmmode \langle\sigma v\rangle R^{2}\else $\langle\sigma
  v\rangle                                                                                    
  R^{2}$\fi}
\def\Dxx{\ifmmode D_{xx} \else $D_{xx}$\fi}
\def\Dpp{\ifmmode D_{pp} \else $D_{pp}$\fi}
\def\ddp{\ifmmode \frac{\partial}{\partial p} \else
  $\frac{\partial}{\partial p}$\fi}
\def\alle{\ifmmode (e^{+}+e^{-}) \else $(e^{+}+e^{-})$ \fi}
\def\pamr{\ifmmode e^{+}/(e^{+}+e^{-}) \else $e^{+}/(e^{+}+e^{-})$
  \fi}
\def\pbarp{\ifmmode \bar{p}/p \else $\bar{p}/p$ \fi}

\def\boverc{\ifmmode B/C \else $B/C$ \fi}
\def\chisq{\ifmmode \chi^2 \else $\chi^2$ \fi}
\def\rchisq{ \ifmmode \chi^2/\dof \else $\chi^2/\mathrm{dof}$ \fi}

\def\sir{ \ifmmode \sigma_{SI,p}\!\times\! R \else $\sigma_{SI,p}\!\times\! R$ \fi}
\def\sdr{ \ifmmode \sigma_{SD,p}\! \times \! R \else $\sigma_{SD,p}\! \times\! R$ \fi}
\def\Olsp{ \ifmmode  \Omega h^2|_{\rm LSP} \else $$ \Omega h^2|_{\rm LSP} \fi}
\def\Owmap{ \ifmmode  \Omega h^2|_{\rm WMAP} \else $\Omega h^2|_{\rm WMAP}$ \fi}
\def\capr{ \ifmmode  C_c \else $C_c$ \fi}

\renewcommand{\thefootnote}{\fnsymbol{footnote}}

\begin{document} \begin{titlepage}
\rightline{\vbox{\halign{&#\hfil\cr
&SLAC-PUB-14390\cr
&SU-ITP-11/06\cr
}}}
\vspace{0.5in}
\begin{center}

{\Large\bf                                                                                                                                        
pMSSM Dark Matter Searches on Ice\footnote{Work supported by the Department of                                                                  
Energy, Contract DE-AC02-76SF00515}}
\medskip

{\normalsize
R.C. Cotta$^{(a)}$, K.T.K. Howe$^{(a,b)}$, J.L. Hewett$^{(a)}$ and T.G. Rizzo$^{(a)}$} \\
\vskip .3cm
$^{(a)}$SLAC National Accelerator Laboratory, 2575 Sand Hill Rd., Menlo
Park, CA 94025 USA \\
$^{(b)}$Physics Dept, SITP, 382 Via Pueblo Mall, Varian Lab
Stanford University, Stanford CA, 94305-4060
\vskip .3cm
\end{center}

\begin{abstract}
We explore the capability of the IceCube/Deepcore array to discover signal 
neutrinos resulting from the annihilations of Supersymmetric WIMPS
that may be captured in the solar core.  In this
analysis, we use a previously generated set of $\sim 70$k model points
in the 19-dimensional parameter space of the pMSSM which satisfy existing 
experimental and theoretical constraints.  Our calculations employ
a realistic estimate of the IceCube/DeepCore effective area that has been
modeled by the IceCube collaboration.  We find that
a large fraction of the pMSSM models are shown to have significant signal 
rates
in the anticipated IceCube/DeepCore 1825 day dataset, including some 
prospects for an early discovery.  Many models where the LSP only constitutes a
small fraction of the total dark matter relic density are found to have 
observable
rates.  We investigate in detail the dependence of the signal neutrino fluxes 
on the LSP mass, weak eigenstate composition, annihilation products and
thermal relic density, as well as on the spin-independent and spin-dependent
scattering cross sections.  Lastly, We compare the model coverage of
IceCube/DeepCore to that obtainable in near-future direct detection
experiments and to pMSSM searches at the 7 TeV LHC.
\end{abstract}


\end{titlepage}

\renewcommand{\thefootnote}{\arabic{footnote}}

  \section{Introduction}  
  \label{sec:intro}

The nature of Dark Matter (DM), which makes up $\sim 25\%$ of the energy budget of the universe \cite{Bertone:2004pz,Primack:1988zm,turner} is one of the greatest mysteries of modern physics. That DM is `dark' implies that it is electrically neutral and thus has only been probed up to now through its gravitational interactions. Although there are many candidate particle physics scenarios which predict various kinds of DM \cite{Steffen:2007sp}, perhaps the most attractive possibility is that the bulk of the DM takes the form of a thermally produced, weakly interacting massive particle (WIMP) \cite{Feng:2010gw}. In most WIMP scenarios that also address the gauge hierarchy problem, the DM state is just the lightest, colorless and neutral member of an entire family of new particles; the lightest neutralino of R-Parity conserving supersymmetry (SUSY) \cite{mssmrev} is arguably the most popular realization for DM in these kinds of models. 

Of course, at the end of the day, only detailed experimental measurements will be able to resolve the issue of DM's nature and data from multiple channels will be necessary. If DM is indeed a WIMP in a scenario such as described above, it could be discovered in the cascade decays of new, but heavier, colored states with TeV-scale masses which are produced with large cross sections at the LHC, with the DM appearing in the detectors as missing transverse energy (MET) \cite{Aad:2009wy,Ball:2007zza}. Similarly, the DM particle may scatter off of a nucleus, depositing energy in an underground detector and be directly observed via the resulting nuclear recoil \cite{Baudis:2007dq,Gaitskell:2004gd}. Finally, there are a number of ways that a WIMP may be indirectly observed through astrophysics experiments. For example, WIMP annihilation in the galactic center and/or halo, or in other galaxies, may be observed through the resulting annihilation products such as photons, positrons, antiprotons in satellite or ground-based observatories \cite{Bertone:2004pz,Porter:2011nv}. A variant of this possibility is to observe the neutrinos that result from the annihilation of WIMPS that have been gravitationally captured in the core of the sun as it sweeps through the DM halo in the galaxy. This is the possibility that we will discuss below. 

In this paper we will explore the capability of the IceCube detector \cite{DeYoung}, in conjunction with its densely instrumented sub-detector DeepCore \cite{deepcore}, which are installed in the ice near the South Pole, to detect neutrinos that would result from neutralino pair annihilation in the solar core. We derive our solar DM signals from models within the phenomenological Minimal Supersymmetric Standard Model (pMSSM), which is a 19-parameter version of the more general MSSM \cite{Berger:2008cq}. Specifically, we will examine the anticipated IceCube signal rates for a large number of points, $\sim 70$k, within this 19-dimensional space that lead to sparticle properties that are consistent with all of the existing experimental and observational constraints. We can then compare the capability of Ice Cube in exploring this 19-dimensional parameter space to that of other direct and indirect DM searches as well as to the ability of the LHC to find SUSY signatures within this same general framework.  

Why should one study such a large supersymmetric parameter space? General soft SUSY breaking within the MSSM leads to a scenario with over $\sim 100$ a priori free parameters so that it is impossible to study in full detail. One approach in dealing with this problem is to consider specific, well-motivated SUSY breaking scenarios, such as mSUGRA \cite{msugrab}, GMSB \cite {gmsbb} or AMSB \cite {amsbb}, all of which lead to a drastic reduction in the number of free parameters (to only $\sim 3-5$). Detailed studies of the resulting parameter space can then be easily performed. One problem with such studies is that they may bias us as to the nature of SUSY signals when performing DM (or collider) searches. An alternative to these top-down methods is to be less prejudicial and to instead follow a bottom-up approach such as that we have employed in a number of recent analyses \cite{Conley:2010du,Cotta:2010ej,Cotta:2009zu}, and will make further use of here. By imposing a set of theoretically and experimentally well-motivated constraints on the general MSSM, without making any reference to the specific mechanism of SUSY breaking, we are able to reduce the $\sim 100$ dimensional parameter space to one with `only' 19 parameters, \ie, the pMSSM, which is significantly more manageable. Such an approach has the advantage of being far more general than any specific SUSY breaking scenario and allows one to be in some sense agnostic about the SUSY mass spectrum, uncovering regions of parameter space that lead to distinct model characteristics and experimental signatures. 

There have been many previous studies of supersymmetric predictions for IceCube/\linebreak DeepCore (IC/DC), employing various strategies for simplifying the full MSSM parameter space. Detailed analyses of specific benchmark mSUGRA models were performed in \cite{Barger:2011em}. In the work \cite{Wikstrom:2009kw} predictions for points in 7- and 9- dimensional subspaces of the MSSM were investigated, while CMSSM predictions for IC/DC have been investigated in \cite{Ellis:2009ka} and in a bayesian framework in \cite{Trotta:2009gr}. Recent considerations of the IC/DC dark matter search outside of the context of the MSSM can be found in \cite{Fukushima:2011df}\cite{Barger:2010ng}. In this work, we expand upon these previous studies and examine the capabilities of IC/DC to detect WIMP signatures from a large set of SUSY models ($\sim 70$k) from the broad 19-dimensional parameter space of the pMSSM. We will find that a large fraction of these models have significant signal rates in the anticipated IC/DC 1825 day dataset. In some cases, the rate is large enough for early discovery. We find that observable rates are expected even for models where the LSP only constitutes a small fraction of the total dark matter relic density. In performing these analyses, we employ a realistic estimate of the IC/DC effective area as modeled by the IC/DC collaboration. In addition, we also examine in detail the dependence of the signal neutrino fluxes on the LSP mass, weak eigenstate composition, annihilation products and thermal relic density, as well as on the spin-independent and spin-dependent scattering cross sections. Coverage of the pMSSM model sample at IC/DC is compared to that obtainable in near-future direct detection
experiments and to pMSSM searches at the 7 TeV LHC.

In the next Section we will briefly discuss the techniques we employed in the generation of the SUSY models corresponding to the $\sim 70$k pMSSM parameter space points we consider in this analysis and the various constraints that were imposed on their selection. In Section 3, we present our analysis method with the final results presented in Section 4. A discussion of these results and our conclusions will then follow in Section 5.  

  \section{Generation of the pMSSM Model Set}
  \label{sec:pMSSM}

In this paper, we explore the sensitivy of the IceCube/Deep Core
array in a broad region of Supersymmetric parameter space.
This is similar in spirit to
Ref. \cite{Cotta:2010ej}, which examined the indirect detection
of dark matter in the pMSSM from its annihilation into electron-positron pairs.
The set of pMSSM models that we investigate 
was generated previously in Ref. \cite{Berger:2008cq} and totals approximately
$\sim 70$k points in the 19-dimensional pMSSM parameter space.  Hereafter we
refer to a point in this parameter space as a model. 
In this Section, we briefly review the procedure employed to generate this model sample.  

We study the 19-dimensional parameter space of the pMSSM \cite{Djouadi:2002ze}.
This set of parameters results from imposing the following minimal set of assumptions 
on the general R-Parity
conserving MSSM: ($i$) the soft parameters are taken to be real so that
there are no new CP-violating sources beyond those in the CKM matrix; 
($ii$) Minimal Flavor Violation \cite{D'Ambrosio:2002ex} is taken to be
valid at the TeV scale; ($iii$) the first two generations of sfermions with the same quantum numbers are taken to be degenerate and to have
negligible Yukawa couplings and ($iv$) the Lighteset Supersymmetric Particle (LSP)
is taken to be the lightest neutralino which is assumed to be 
a stable thermal WIMP. No assumptions about the physics at high scales or SUSY breaking
mechanisms were employed.
The first three of these conditions are applied in order to avoid issues 
associated with constraints from the flavor sector.  These assumptions reduce
the SUSY parameter space to 19 free soft-breaking parameters which are given by the three gaugino masses, $M_{i=1-3}$, ten sfermion
masses $m_{\tilde Q_1,\tilde Q_3,\tilde u_1,\tilde d_1,\tilde u_3,\tilde d_3,
\tilde L_1,\tilde L_3,\tilde e_1,\tilde e_3}$, 
the three $A$-terms associated with the third generation  ($A_{b,t,\tau}$), and
the usual Higgs sector parameters $\mu$, $M_A$ and $\tan \beta$.

To produce the set of pMSSM parameter points used in this paper, 
we performed numerical scans over
the space formed by these 19 parameters. This required a selection
of the parameter range intervals as well as an assumption about the nature of the scan prior for how the points were chosen within these intervals. These
issues are described in detail in our previous works \cite{Berger:2008cq,Conley:2010du,Cotta:2010ej}. Here, we simply note
that two scans were performed: one employing a flat prior beginning with $10^7$ points, and
a second with a log prior employing $2\times 10^6$ points. The main
distinctions between these two scans directly relevant to our analysis here are that ($i$)  all SUSY mass parameters were
restricted to be $\leq 1$ TeV for the flat prior case, while for the log sample the upper limit on mass parameters was raised to 3 TeV, and ($ii$) the
choice of the log prior generally
leads to a more compressed sparticle spectrum than does the flat prior case.  Note that the
restriction on the upper limit for the mass parameters ensures relatively large production cross sections at the LHC for the case of the flat prior model sample.  Most of the
results presented below were obtained with the larger flat prior model set.

Once these points were generated, we subjected them to a large set of theoretical 
and experimental constraints, and required consistency.  This ensures that the model
sets are valid to study.    We briefly review this set of restrictions
here{\footnote {For full details, see Ref. \cite{Berger:2008cq}}}:
($i$) The theoretical constraints we imposed are that the spectrum was required to 
be tachyon free, color and charge breaking minima 
must be avoided, a bounded Higgs potential must be obtained and electroweak symmetry breaking must be consistent. ($ii$) We employed a number of constraints from the flavor
sector and precision electroweak data 
arising from the measurements of $(g-2)_\mu$, $b\to s\gamma$, $B\to \tau \nu$, $B_S \to \mu^+\mu^-$, meson--anti-meson mixing, the invisible width of the $Z$ and
$\Delta \rho$. ($iii$) We required that the LSP contribution to the dark matter relic
density not exceed the upper bound determined by WMAP.  Note that we did not require
the LSP to saturate the measured relic density; this leaves room for the existence of 
other dark matter candidates.  In addition, limits 
from dark matter direct detection searches were also applied. ($iv$) The restrictions  resulting from the numerous direct searches at LEP for both
the SUSY particles themselves as well as the extended SUSY Higgs sector  were
imposed. Here, some care was necessary as some of these searches needed to be
re-evaluated in detail due to SUSY model-dependent assumptions present in the analysis which we removed. ($v$) Finally, the null results from the set
of Tevatron SUSY sparticle and Higgs searches were imposed. 
The most restrictive data were found to be those from searches for stable charged
particles \cite{Abazov:2008qu} and those looking for an excess of multijet plus 
MET events \cite{:2007ww}.
We note that in the latter case, the search strategies were designed for kinematics expected in mSUGRA-inspired models.  We thus simulated the search
in some detail, at the level of fast Monte Carlo, for our full model sample.

After this set of constraints was imposed,
$\sim 68.4k$ models from the flat prior set survived this analysis chain, 
as well as a corresponding set of $\sim 2.9k$ models from the log prior sample. 
This forms the set of models that we will consider in our following analysis of the IceCube/DeepCore capabilities to
detect dark matter.

  \section{Solar Neutrino Rate Calculations}
  \label{sec:nurates}

\indent If the local DM halo is composed of WIMP dark matter with an empirically estimated local density $\rho_0 \sim 0.3 \rm{~GeV/cm^3}$ we may be able to observe its presence as the result of WIMP scattering interactions in the sun. As the sun passes through the DM halo, WIMPs can scatter off of solar nuclei into orbits that are bound in the gravitational potential of the sun, eventually settling to the core of the sun after repeated scatterings \cite{Press:1985ug}. Subsequent annihilations of captured WIMPs generically produce energetic particles, including neutrinos that can propagate out of the sun to terrestrial neutrino detectors. 

The instantaneous number of WIMPs in this captured population, $N(t)$, can be modeled as
\begin{equation}
 	\frac{dN}{dt} = C_c-C_a N(t)^2,
	     \label{solarwimps}
\end{equation}
where $C_c$ is the capture rate and $C_a$ is proportional to the WIMP thermal annihilation cross section $\sigv$ (the constant of proportionality is the effective volume for WIMP scattering in the sun). The annihilation rate of captured WIMPs can be written as $\Gamma_a = C_a N^2$. The contribution to the total capture rate $\capr$ from a shell of solar volume $V$ at radius $r$ is, schematically,
\begin{equation}
 	\frac{dC_c}{dV} = \int dv f(v,r) \sum_i \left(\sigma_i(v) \frac{\rho_\chi}{M_\chi}v\right)n_i(r) P_i(v,r).
	     \label{captureeq}
\end{equation}
The sum runs over the elements present in the sun. The number density of a specific target element is $n_i(r)$ and the associated cross-section for WIMP-nuclei elastic scattering on target nuclei of this type is $\sigma_i$. The probability of a scatter resulting in a bound orbit is $P_i(v,r)$. The local sun frame WIMP velocity distribution is $f(v,r)$ and $\rho_\chi$ is the local halo density of the scattering WIMP species\footnote[1]{$\rho_\chi$ may or may not equal the empirically estimated total DM density $\rho_0 \sim 0.3 \rm{~GeV/cm^3}$, as we describe in the next subsection LSP neutralinos in our pMSSM model set have $\rho_\chi\leq\rho_0$, and the resulting neutrino signals must be appropriately scaled for each model.}. Analytic expressions for these contributions can be found in \cite{Gould:1987ir}. For the sun, the dominant contribution to the capture rate is typically from spin-dependent elastic scattering of WIMPs off of hydrogen nuclei, $\sigma_{SD,p}$. Spin-independent elastic scattering off of heavier elements, though rare because of their low estimated solar abundances, may also provide important contributions due to the $A^2$ coherent enhancement of spin-independent scattering (oxygen, helium and neon may be most important in this regard \cite{Agrawal:2010ax}).  

Equation \ref{solarwimps} can be easily solved; the annihilation rate of captured WIMPs at present time is given by \cite{Wikstrom:2009kw}
\begin{equation}
 \Gamma_a \equiv \frac{1}{2}C_aN^2(\tau_\odot)= \frac{C_c}{2}\tanh^2\frac{\tau_\odot}{\tau_{\rm eq}},
 \label{capvanneq}
\end{equation}
where $\tau_{\rm eq}=(C_a C_c)^{-\frac{1}{2}}$ is the time required for WIMPs to reach equilibrium and $\tau_\odot = 4.5\cdot10^9\;\rm{yr}$ is the age of the sun. For $\tau_\odot\gg\tau_{\rm eq}$ the solar WIMP population is in equilibrium, $dN/dt =0$, and the annihilation rate approaches half the capture rate, $2\Gamma_a/C_c\approx1$. In this limit the annihilation rate has no explicit dependence on $\langle\sigma v\rangle$ and is instead completely determined by the capture rate, $C_c$. The condition of equilibrium is commonly attained for the combinations of $\sigv$, $\sigma_{SD,p}$ and $\sigma_{SI,p}$ that are usually found for MSSM models, but is not completely general. This will be demonstrated in the more detailed discussion of the calculation of neutrino rates from our pMSSM models, to which we now turn. 

\subsection{Solar Neutrino Signals from the pMSSM}
  \label{sec:pmssmrates}

\indent We now remark on some specific aspects of this calculation as they relate to our pMSSM model results and their interpretation. 

The magnitude and energy spectrum of the signal neutrino flux will depend on the properties of not only the LSP neutralino $\LSP$, but to some extent on substantially all of the masses and couplings needed to describe a particular pMSSM model. In order to calculate signal neutrino rates from each of our $\sim$ 71k pMSSM models we rely on the computational package DarkSUSY 5.0.5 \cite{Gondolo:2004sc}. For each of our models we input SUSY Les Houches Accord files via SLHAlib \cite{slha} to DarkSUSY, which subsequently calculates annihilation and scattering cross-sections. DarkSUSY uses the analytic formulae in \cite{Gould:1987ir} (as described in Ref. \cite{Wikstrom:2009kw}) to derive the rate and spectra of signal neutrinos injected at the solar core by the solar $\LSP$ population. These injection spectra are then propagated out of the sun and to/through the earth to the detector via the package WimpSim \cite{Blennow:2007tw}, which is embedded in DarkSUSY 5.0.5. 

It is important to note that, because we have employed the WMAP measurement of the DM relic density only as an upper bound on selecting pMSSM models, we must appropriately rescale the empirical estimate $\rho_0 = 0.3 \rm{~GeV/cm^3}$ for the total local DM energy density by the factor
\begin{equation}
 R=\frac{\Omega_{\LSP}}{\Omega_{\rm{WMAP}}},
     \label{OmegaR}
\end{equation}
where $\Omega_{\rm{WMAP}}h^2=0.1143$ \cite{Komatsu:2008hk}. For definiteness we identify the set of pMSSM models which yield $\Omega_{\rm{\LSP}}h^2>0.10$ ($R>0.875$) as models which saturate the WMAP bound and distinguish this subset of models in the figures that follow. LSPs in models with $R$ substantially less than 1 are interpreted as comprising one component of a multi-component DM halo. 

As regards the local WIMP velocity distribution $f(v,r)$, our calculations employ a standard Maxwellian distribution with dispersion $\overline{v} = 270 \rm{~km/s}$ (in the halo frame), boosted into the sun's frame ($v_\odot=220 \rm{~km/s}$) and modified locally by the solar potential. It is difficult to quantify the uncertainty in the WIMP velocity distribution, but choosing a different form for the velocity distribution would likely yield at most an $\mathcal{O}$(1) change in the capture rate (see $\eg$ \cite{Wikstrom:2009kw}).

As mentioned in the previous section, solar WIMP capture-annihilation equilibrium has been seen to occur in most, but not all, of the previously studied MSSM models. We define capture-annihilation equilibrium for our models via the criterion: 
\begin{equation}
 \frac{2\Gamma_a}{C_c} > 0.9.
 \label{equilibrium}
\end{equation}

By this definition, 6.5\% of our flat-prior and 14\% of our log-prior pMSSM models are found to be out-of-equilibrium. Though out-of-equilibrium models typically lead to low solar neutrino rates we find a wide variety of predictions for out-of-equilibrium models in our set, and we will also distinguish this subset of models in the figures that describe our results.

The numerical calculation of elastic scattering cross sections also deserves further comment. The calculation of $\LSP$-nuclei cross sections involves first the parton level calculation of $\LSP$ scattering off of the nucleonic constituents. Several different analytical treatments and associated numerical implementations of the parton-level scattering calculations are available in DarkSUSY and the choice for this analysis is important because of the sheer phenomenological variety of the sparticle spectra in our pMSSM model set. We employ two non-default options: \emph{(i)} to include neutralino and quark mass terms along with the squark mass terms in the denominators of propagators, and \emph{(ii)} to include the DarkSUSY implementation of the extra twist-2 and box diagram contributions to the spin-independent cross section that were originally derived in \cite{Drees:1993bu}. The approximation to leave out these contributions is valid when squark masses are much greater than the quark and neutralino masses, which is not the case for generic models in our set. Overall, the default calculation was seen to diverge from more accurate calculations by more than an order of magnitude for $\sim$ 20\% of the models in our flat-prior set, and by up to five orders of magnitude for nearly pure bino-like LSPs which satisfied the WMAP relic density bound via large squark co-annihilations. Implementation of the non-default options addressed these issues.

There are, of course, uncertainties that are also incurred in the estimates of the nuclear matrix elements that are necessary for going from the parton level cross-sections to the $\LSP$-nucleon cross sections (for recent accounts see \cite{formfactors}). These ambiguities have been shown to translate to an uncertainty in the capture rate of $\sim10\%$ for dominantly spin-dependent contributions and by an $O(1)$ factor for dominantly spin-independent contributions \cite{Ellis:2009ka}. An additional $\sim 20\%$ error also applies to the latter case depending on the choice of solar models with differing metallicities \cite{Ellis:2009ka}. In this work we use the following nuclear form factors (in the language of, $\eg$, \cite{Belanger:2008sj}):

\begin{eqnarray}
f^p_u=0.023,\hspace{10 mm} f^p_d&=&0.033,\hspace{10 mm} f^p_u=0.26, \nonumber \\
f^n_u=0.018,\hspace{10 mm} f^n_d&=&0.042,\hspace{10 mm} f^n_u=0.26,\\
\Delta^p_u=0.842,\hspace{9 mm}\Delta^p_d&=&-0.427,\hspace{6 mm}\Delta^p_s=-0.085. \nonumber 
 \label{formfactorvalues}
\end{eqnarray}

Other sources of uncertainty in the calculation include the modeling of the effects of charged and neutral current interactions as well as oscillations that neutrinos undergo in their propagation through the sun, to the earth and through the earth. We use the default DarkSUSY/WimpSim settings for solar composition, and neutrino propagation/oscillation parameters. The effects of varying these parameters on the resulting signal flux have been investigated in \cite{Blennow:2007tw}, where it was found that significant uncertainties ($\sim$ factors of a few) may be present in the treatment of oscillations, particularly from our lack of knowledge of $\theta_{13}$ and of the neutrino mass hierarchy. Accurate accounts of these errors will, of course, be important in determining the significance of any experimental result. However, as we are here focused on the SUSY model dependence of our predictions, we take into account the sources of error that we estimate to be the largest in this regard.

\subsection{The IceCube/DeepCore Solar WIMP Search}
  \label{sec:icdcsearch}

\indent The IceCube neutrino telescope observes the Cerenkov light emitted from charged leptons as they transit through the Antarctic ice. The DeepCore extension of IceCube is a densely instrumented subsection of the larger IceCube detector, which is situated in the deepest ice near the center of the IceCube array \cite{deepcore}. This setup, using the larger IceCube detector as a veto for the embedded DeepCore detector, allows for a reduced neutrino energy threshold and hence better performance in searching for the neutrino signal resulting from WIMP annihilation. As neutrinos from WIMP annihilation pass through the ice and earth in and below the detector they produce detectable muons via charged current scattering which can then be identified and measured by the instrument. There are three backgrounds for the signal neutrinos produced by WIMP annihilation in the sun: \emph{(i)} atmospheric muons, \emph{(ii)} atmospheric muon neutrinos, and \emph{(iii)} neutrinos from cosmic ray interactions in the solar atmosphere. A large potential background from muons produced in cosmic ray interactions in the atmosphere is essentially eliminated by requiring candidate signal muons that come from the northern hemisphere, and hence from the conversion of muon-neutrinos that are capable of transit through a large portion of the earth. Because of this, the search that we describe here requires 1825 days ($\sim5 \:\rm{yr}$) of useable data, but must be taken over a $\sim10\:\rm{yr}$ timescale.

The number of observed signal events depends on the energy- and angular-dependent \emph{effective area} of the detector. The effective area is determined by the IC/DC collaboration via monte carlo simulation. Here an initially generated population of neutrinos of various energies, incident on the earth from various directions, are transported through the earth, converted to muons, and transported to the detector, where, finally, a number of low-level data quality criteria are required in order to accept the event. In this way an energy- and angular-dependent efficiency is derived ($\ie$, taking the ratio of accepted events to generated neutrinos) which, when multiplied by the fiducial area of the detector, gives an effective area that should be convolved with raw signal neutrino fluxes\footnote{In particular we use the ``SMT8/SMT4" effective areas, derived as a preliminary estimate for the IceCube 80-string+DeepCore configuration by the IC/DC collaboration \cite{IceEffArea}.} in order to calculate the number of observed events. The effective area increases rapidly with increasing incident neutrino energy as both the $\nu_{\mu}-\mu$ conversion cross-section and the range of the produced muons grow approximately linearly with the neutrino energy; signal event rates are thus larger for stiffer signal neutrino spectra. The angular dependence of the effective area is fairly flat for neutrinos incident from the northern hemisphere and it is dramatically suppressed for neutrinos incident from the southern hemisphere, where stringent angular cuts are necessary to reduce the atmospheric muon background. In performing the detector effective area convolution we integrate over the northern hemisphere only and over neutrino energies $>\!10\gev$. The raw neutrino fluxes given by the default DarkSUSY/WimpSim setup represent an average over the incident rates during the period spanning the spring and autumn equinoxes ($\ie$, the period during which the sun is below the horizon for a detector at the South Pole).

We define the detected neutrino rate, $ \Phi^D_{\nu}$ (events/yr), as
\begin{equation}
 \Phi^D_{\nu} = \int \bigg(A^{\nu_\mu}_{\rm{eff}}(E)\frac{d\Phi_{\nu_\mu}}{dE} + A^{\overline{\nu}_\mu}_{\rm{eff}}(E)\frac{d\Phi_{\overline{\nu}_\mu}}{dE}\bigg) dE,
     \label{aeffeq}
\end{equation}
where $d\Phi_{\nu_\mu,\overline{\nu}_\mu}/dE$ are the raw differential neutrino/antineutrino fluxes calculated here using DarkSUSY/WimpSim, $E$ is integrated over the neutrino energies $>\!10\gev$, and $A^{\nu_\mu,\overline{\nu}_\mu}_{\rm{eff}}(E)$ are the energy-dependent neutrino effective areas appropriate for neutrinos and antineutrinos, respectively.

This effective area does not include cuts which are needed to further reduce the atmospheric muon background and may lead to a reduction of the effective areas used here by a factor of $\sim5$ \cite{Wikstrom:2009zz}. To distinguish the annihilation signal from the isotropic atmospheric muon neutrino background, the IC/DC collaboration employs a directional search that looks for a statistical excess correlated with the solar angle.  Ref. \cite{Wikstrom:2009zz} estimates that the planned 1825 day dataset will have a $90\%$ exclusion sensitivity for a signal event rate of $\sim 8~\rm{events/yr}$. Taking into account the factor of five signal reduction by implementing directional cuts, we estimate the $90\%$ confidence exclusion limit for SMT8/SMT4 trigger level signal rates at $\Phi^D_{\nu} \sim 40~\rm{events/yr}$. From this we obtain a discovery threshold for trigger level signal events of $\Phi^D_{\nu} \sim 90-100~\rm{events/yr}$. In the work \cite{Trotta:2009gr}, it is estimated that 11.5 events/yr is the appropriate level for 90\% exclusion by the Feldman-Cousins construction \cite{Feldman:1997qc}, and that 31.6 events/yr is appropriate for discovery with $\rm{S}/\sqrt{B}\geq5$. The range $\sim10-100$ events/yr thus constitutes a plausible estimate for exclusion/discovery criteria (where at the lower end one approaches the irreducible directional background of $\mathcal{O}(10)$ trigger level events per year that are estimated from neutrinos produced in cosmic ray interactions in the solar atmosphere \cite{Seckel1991}\cite{Wikstrom:2009zz}). More accurate criteria await a more detailed estimate from within the IceCube/DeepCore Collaboration. \emph{Throughout the rest of this paper we will use $\Phi^D_{\nu}=40~\rm{events/yr}$ as a criterion for \textbf{exclusion}, and when comparing to the discovery potential of LHC searches, we will use $\Phi^D_{\nu}=100~\rm{events/yr}$ as a conservative criterion for \textbf{discovery}}. We note that the qualitative conclusions presented in the sections that follow are fairly insensitive to this choice, though the overall number of models that may be visible in our set varies considerably with the choice of discovery criteria.

We note that the IceCube collaboration has also computed expected exclusion limits on WIMP-nucleon elastic scattering cross sections that may be attained in the 1825 day IC/DC search \cite{Braun:2009fr}. These limits are based on the conversion factors developed in \cite{Wikstrom:2009kw} and assume that \emph{(i)} the WIMP is in solar capture/annihilation equilibrium, \emph{(ii)} the WIMPs make up the entirety of the local dark matter density and \emph{(iii)} the WIMPs annihilate exclusively into hard channels ($\ie$, $W^{+}W^{-}$, $Z^{0}Z^{0}$ and $\tau$-pairs). We will compare this result with those obtained for the models in our set in Section \ref{sec:excomps}. We note here that the authors of \cite{Wikstrom:2009kw} discussed that differences in annihilation final state channels could generate about an order of magnitude variance in the predictions of models with fixed $C_c$ and $m_{\LSP}$, and that, in addition, about one order of magnitude variance could be attributed to varying the LSP mass with fixed $C_c$ and annihilation final states (both effects are primarily due to the highly energy-dependent detector effective area). This finding is largely echoed in our analysis. Since all of the other uncertainties that we have discussed are $\mathcal{O}(1)$, we believe that our treatment of the detector effective area should allow us to take directly into account the dominant source of approximation, while allowing a more accurate interpretation of the SUSY model dependence of our results.

  \section{Results from the pMSSM Model Set}
  \label{sec:results}

  \indent In this section we describe the detected neutrino flux rates that can be expected in IceCube/DeepCore (IC/DC) for the models in our pMSSM set. We investigate the dependence of the resulting flux on relevant SUSY model parameters: $m_{\LSP}$, $\sigma_{SI,p}$, $\sigma_{SD,p}$, $\Omega h^2|_{\LSP}$ and the annihilation rates into various SM final states. We compare the results obtained from models in our flat-prior set with those in our log-prior set. We combine estimates of background event rates with signal flux predictions to estimate which of our pMSSM models may be detected or ruled out by IC/DC in the 1825 day solar WIMP search data. Finally, we discuss the ability of IC/DC relative to current/planned direct detection experiments, as well as the LHC, to discover or constrain supersymmetric DM.
   
\subsection{Basic Results}
\label{sec:basic} 
 
\begin{figure}[hbtp]
    \centering
    \includegraphics[width=1.0\textwidth]{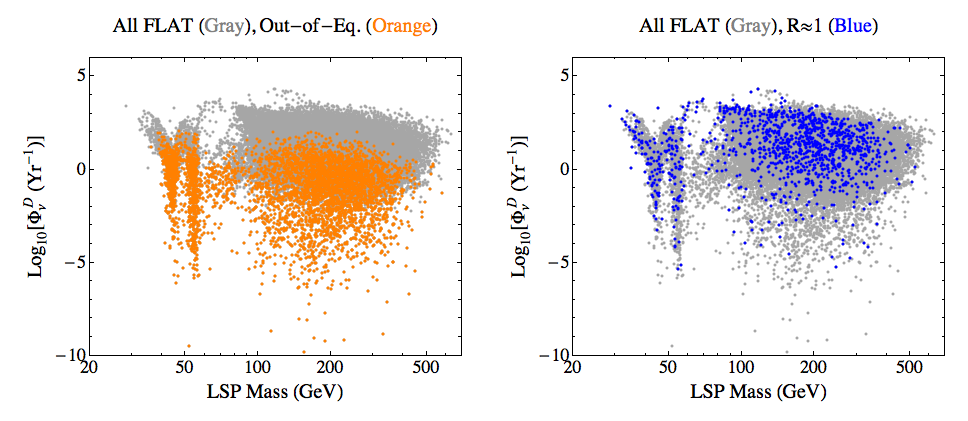}
    \caption{We display points representing models in our flat-prior pMSSM model set in the detected neutrino flux, $\Phi^D_\nu$ (see Eqn.~(\ref{aeffeq})), vs.\ LSP mass plane. Grey points represent all of the models in this set while orange points denote pMSSM models which are out of capture/annihilation equilibrium according to Eqn.\ (\ref{equilibrium}) and blue points represent models whose LSPs form substantially all of dark matter, $\Omega_{\rm{\LSP}}h^2>0.10$ ($R\approx1$).}
    \label{figs:Fmass}
  \end{figure}

\begin{figure}[hbtp]
    \centering
    \includegraphics[width=1.0\textwidth]{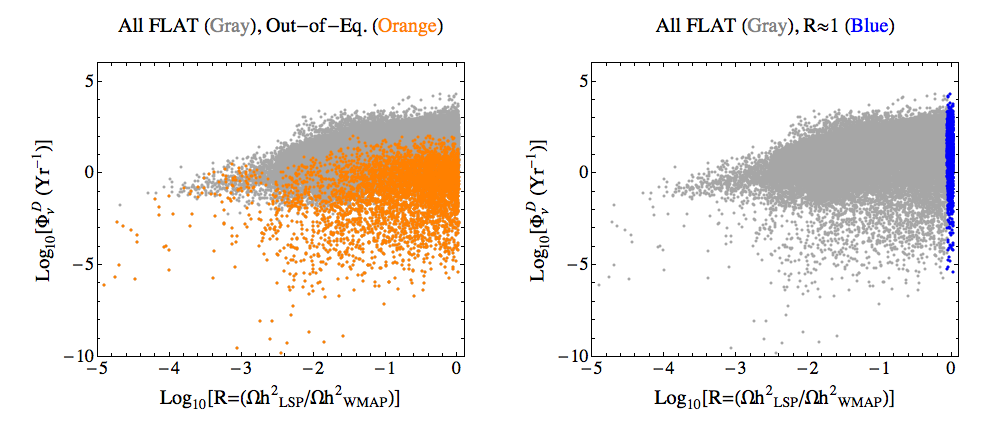}
    \caption{We display points representing models in our flat-prior pMSSM model set in the detected neutrino flux, $\Phi^D_\nu$ (see Eqn.~(\ref{aeffeq})), vs.\ $R$ (see Eqn.~(\ref{OmegaR})) plane. Grey points represent all of the models in this set, orange points denote pMSSM models which are out of capture/annihilation equilibrium according to Eqn.\ (\ref{equilibrium}) and blue points represent models whose LSPs form substantially all of dark matter, $\Omega_{\rm{\LSP}}h^2>0.10$ ($R\approx1$). $R\approx1$ models are, of course, situated at the very right-hand edge this figure.}
    \label{figs:Fr}
  \end{figure}

\begin{figure}[hbtp]
    \centering
    \includegraphics[width=1.0\textwidth]{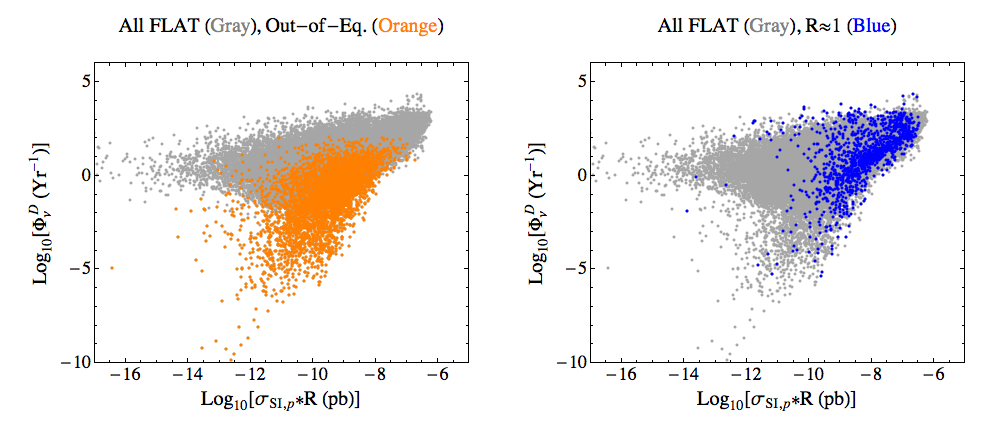}
    \caption{We display points representing models in our flat-prior pMSSM model set in the detected neutrino flux, $\Phi^D_\nu$ (see Eqn.~(\ref{aeffeq})), vs.\ $\sir$ plane. Grey points represent all of the models in this set, orange points denote pMSSM models which are out of capture/annihilation equilibrium according to Eqn.\ (\ref{equilibrium}) and blue points represent models whose LSPs form substantially all of dark matter, $\Omega_{\rm{\LSP}}h^2>0.10$ ($R\approx1$). We note that we include all models which satisfied the direct detection limits that were in place \emph{when these models were generated} \cite{Berger:2008cq} (wherein a factor of 4 error was allowed for nuclear form factor uncertainties). We will consider the effect of recent direct detection limits (including the very recent XENON100 limit \cite{Aprile:2011hi}) in Section \ref{sec:excomps}.}
    \label{figs:Fsi}
  \end{figure}

\begin{figure}[hbtp]
    \centering
    \includegraphics[width=1.0\textwidth]{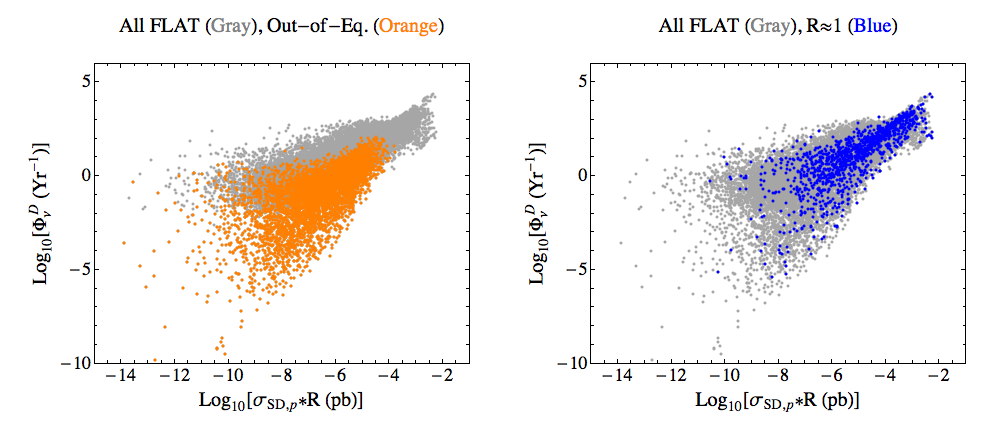}
    \caption{We display points representing models in our flat-prior pMSSM model set in the detected neutrino flux, $\Phi^D_\nu$ (see Eqn.~(\ref{aeffeq})), vs.\ $\sdr$ plane. Grey points represent all of the models in this set, orange points denote pMSSM models which are out of capture/annihilation equilibrium according to Eqn. (\ref{equilibrium}) and blue points represent models whose LSPs form substantially all of dark matter, $\Omega_{\rm{\LSP}}h^2>0.10$ ($R\approx1$). We note that $\sdr$ is the quantity which is most strongly correlated with high detected neutrino rates.}
    \label{figs:Fsd}
  \end{figure}

\begin{figure}[hbtp]
    \centering
    \includegraphics[width=1.0\textwidth]{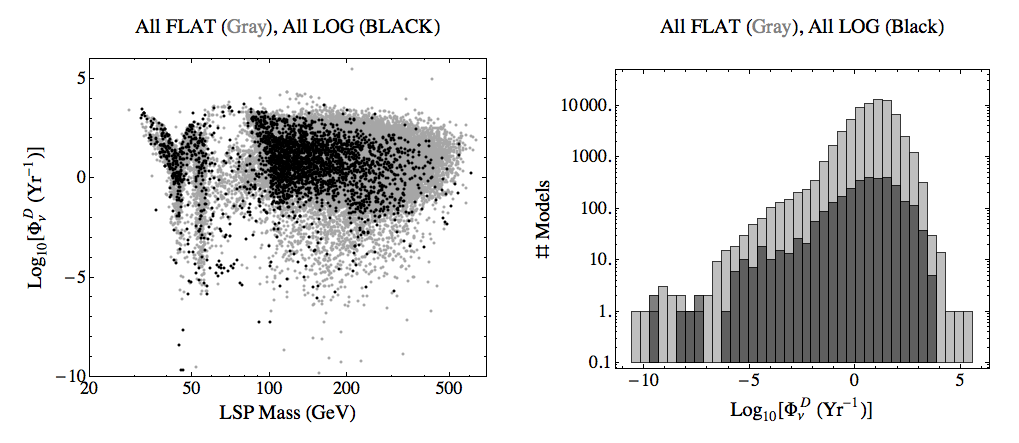}
    \caption{Comparison of the basic results for our flat-prior pMSSM model set with those for our log-prior pMSSM model set. Flat-prior models are represented by grey points while log-prior models are represented by black points. There are $\sim$68.4k models in the flat-prior set and $\sim$2.9k models in the log-prior set. We note the general similarity between these two model samples.}
    \label{figs:flatlog}
  \end{figure}

\indent In Figures \ref{figs:Fmass}-\ref{figs:flatlog} we investigate the dependence of the detected neutrino flux, $\Phi^{D}_\nu$ (the result after convoluting the raw flux spectra with the detector effective area, Eqn.\ (\ref{aeffeq})), on LSP mass, WIMP-proton elastic scattering cross-sections and LSP relic density. In each figure we display points for each of the pMSSM models in our flat-prior model set (grey) or log-prior model set (black). We highlight pMSSM models which are out of solar capture/annihilation equilibrium (as defined in Eqn.\ (\ref{equilibrium})) in orange and pMSSM models that make up substantially all of DM, $\Olsp\approx\Owmap$, in blue. In figures involving the elastic scattering cross sections, we use scaled cross sections, $\eg$, $\sdr$, where $R$ was defined in Eqn.\ (\ref{OmegaR}). Scaled cross-sections are appropriate because the limits placed on elastic scattering cross sections are proportional to the abundance of the WIMP and are usually quoted with the assumption that the scattering particle makes up all of the observed DM\footnote{In using this scaling, we are also assuming a canonical thermal cosmological history and that the DM distributions responsible for the signals are reasonably approximated by the large scale average abundance, $\ie$, that substructure in the DM distribution does not heavily affect the resulting signals.}.
 
One very obvious feature in Figs.\ \ref{figs:Fsi}-\ref{figs:Fsd} is that large elastic scattering cross-sections, especially the spin-dependent WIMP-proton cross-sections, are strongly correlated with large neutrino signal rates. This is to be expected as the normalization of the neutrino spectra from models in capture/annihilation equilibrium is determined by the capture rate, which is in turn determined by the elastic scattering cross-sections. We will turn to a more detailed investigation of the SUSY model dependence of these results in the next section (Section \ref{sec:susymi}).

\begin{figure}[hbtp]
    \centering
    \includegraphics[width=1.0\textwidth]{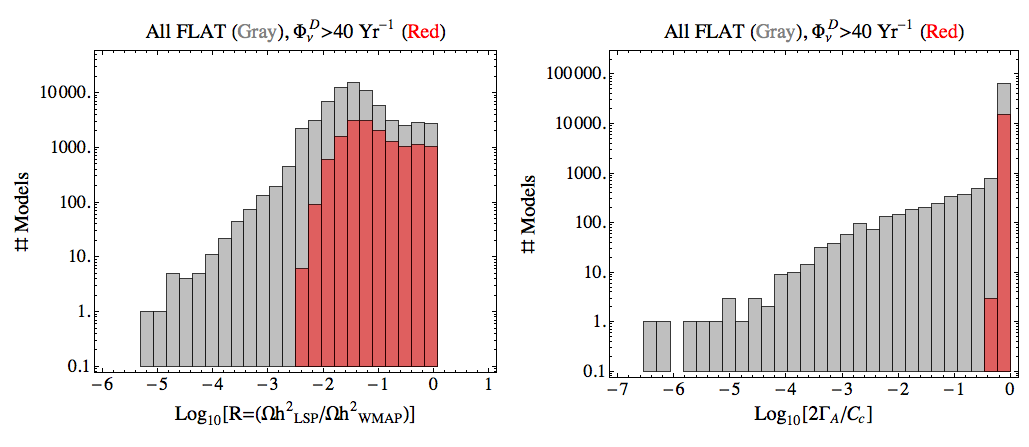}
    \caption{Using the approximate exclusion criterion described in Section \ref{sec:icdcsearch} we compare histograms of models that are expected to be excluded by IC/DC (red) against the full model set (grey) as functions of the LSP relic density and of the ratio $2\Gamma_a/C_c$ (recall Eqn. (\ref{equilibrium})). As described in the text, exclusion performance is not strongly correlated with LSP relic density and we see that IC/DC may be able to discover LSPs with $R\ll1$. In contrast to this we see that essentially none of the models that can be excluded are out-of-equilibrium.}
    \label{figs:hist-disc-r-eq}
  \end{figure}

From Figures \ref{figs:Fmass}-\ref{figs:Fsd} it is difficult to tell how many pMSSM models would have high signal neutrino rates at IC/DC. Figure \ref{figs:flatlog} includes a histogram of the detected neutrino flux rates in our flat-prior model set and we note here that approximately 83\%, 48\%, 8.6\% and 0.6\% of these models are expected to have detected signal rates greater than 1, 10, 100, and $10^3 \: \rm{yr}^{-1}$, respectively. According to the background and signal significance estimates described in the previous section, these results imply that the IC/DC 1825 day dataset may be expected to exclude (discover) $\sim 22\%$ ($8.6\%$) of the models in our flat-prior model set (the $\sim400$ models with signal event rates $>10^3 \: \rm{yr}^{-1}$ would likely be seen with much less than the planned 1825 days of data). Among the subset of pMSSM models whose LSPs make up substantially all of DM the coverage is somewhat better, with $\sim 38\%$ of such flat-prior models being excluded in the 1825 day IC/DC dataset. These results are displayed in Fig. \ref{figs:hist-disc-r-eq}. As expected, pMSSM models which are out of capture/annihilation equilibrium are much more difficult to exclude. Essentially none of the flat-prior out-of-equilibrium model subset are expected to be excluded by our criteria, as is demonstrated in Fig.\ \ref {figs:hist-disc-r-eq}. We also note that capture/annihilation equilibrium is \emph{not} strongly correlated with relic density (this is quite insensitive to our definition of equilibrium), see Fig.\ \ref{figs:Fr}.

In Figure \ref{figs:flatlog} the detected neutrino rates for our \emph{log-prior} pMSSM model set are compared with those for our flat-prior pMSSM model set. By the criterion described above we may expect to exclude about 25\% of the models in our log-prior model set. We note that, besides the difference in statistics, the distributions shown in the histogram Fig. \ref{figs:flatlog} are very similar to each other. While histograms like these are better interpreted as a description of the possibilities within our $\sim\!71\,\rm{k}$ pMSSM model set than as posterior distributions for the MSSM we note the encouraging robustness of these results with respect to a change in scan priors.

From Figures \ref{figs:Fmass}-\ref{figs:Fsd} and Figure \ref{figs:hist-disc-r-eq} it is interesting to note that there is \emph{not} a strong correlation between the relic density of the LSP and the resulting neutrino signal. IC/DC is sensitive to pMSSM models with $R\gsim10^{-2}$ at approximately the same level as those with $R\approx 1$. This is a reflection of the fact that, for models in capture/annihilation equilibrium, the solar WIMP signal scales with relic density\footnote{This simple scaling is of course altered when the $\sigv$ effective at freeze-out is not the same as the effective $\sigv$ in present-day DM halos (for example when co-annihilations are significant \cite{Griest:1990kh}). Many of the models in our set display this property, though it does not significantly change the intuition of these remarks.} as $\Phi^D_\nu\sim(\sdr)\sim1$, for $R\sim1/\sigv$, similarly to the expected scaling in terrestrial direct detection experiments. This is in contrast to the $\tsigv\sim R$ scaling appropriate for the cosmic ray signals from DM annihilation in the galactic halo, galactic center (including, for example, IceCube searches for neutrinos from DM annihilating near the Galactic Center), or other astrophysical DM distributions. A more detailed discussion of the performance of IC/DC as it compares to/complements the LHC, direct detection and indirect detection experiments will follow in Section \ref{sec:excomps}.

\subsection{SUSY Model Dependence}
\label{sec:susymi} 

\indent In order to examine the SUSY model dependence of these results, we first discuss the SUSY model dependence of the \emph{normalization} of the signal neutrino spectra, and then we explore the dependence of the results on the \emph{shape} of the signal neutrino spectra.

\begin{figure}[hbtp]
    \centering
    \includegraphics[width=1.0\textwidth]{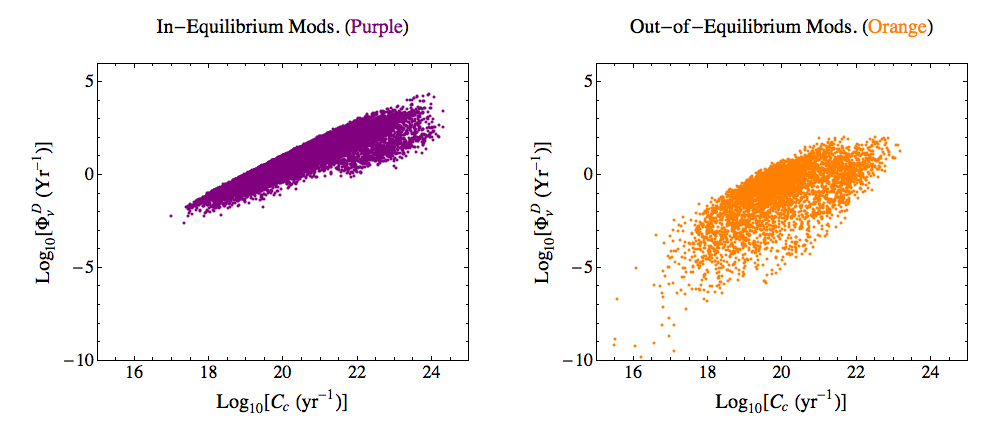}
    \caption{We display points representing models in our flat-prior pMSSM model set in the detected neutrino flux, $\Phi^D_\nu$, vs.\ solar capture rate, $C_c$, plane. Purple points represent models that are \emph{in} capture/annihilation equilibrium according to Eqn. \ref{equilibrium} and orange points denote pMSSM models which are \emph{out} of capture/annihilation equilibrium.}
    \label{figs:dcap}
  \end{figure}

\begin{figure}[hbtp]
   \centering
    \includegraphics[width=1.0\textwidth]{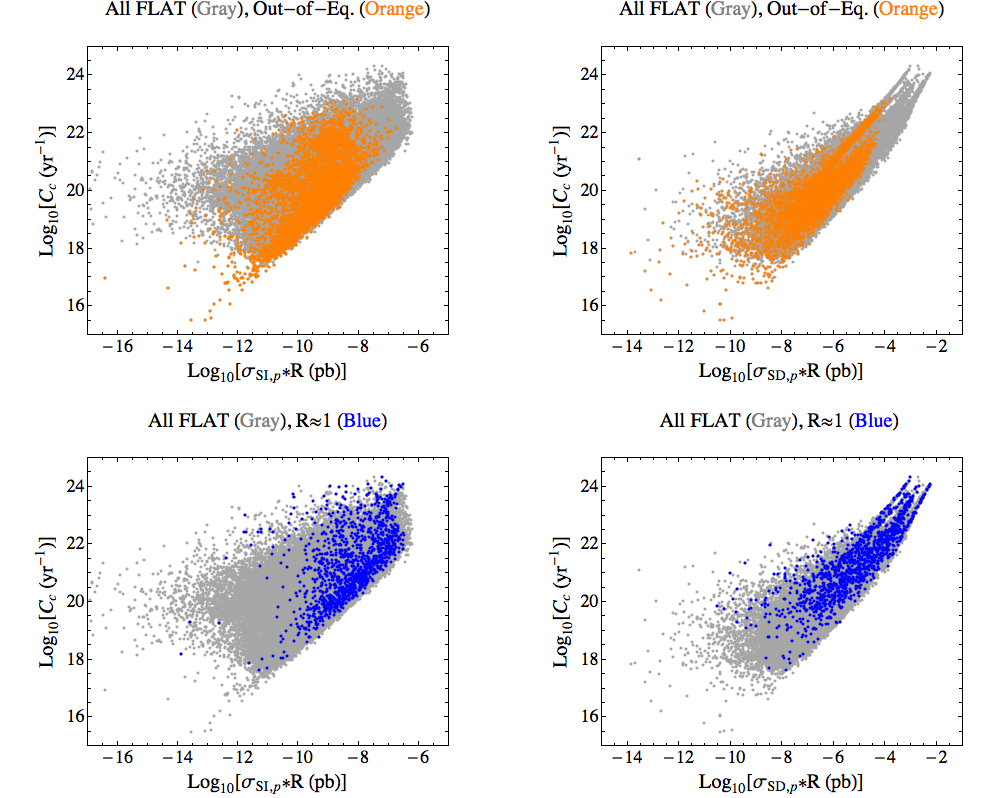}
    \caption{We display points representing models in our flat-prior pMSSM model set in the capture rate, $C_c$, vs.\ elastic scattering, $\sir$ or $\sdr$ planes. Grey points represent all of the models in this set, orange points denote pMSSM models which are out of capture/annihilation equilibrium according to Eqn. (\ref{equilibrium}) and blue points represent models whose LSPs form substantially all of dark matter, $\Omega_{\rm{\LSP}}h^2>0.10$ ($R\approx1$). The strong correlation between $C_c$ and $\sdr$ explains the strong correlation observed between $\Phi^D_\nu$ and $\sdr$ (Fig. \ref{figs:Fsd}). The strong correlation between $C_c$ and $\sdr$ (and relatively weak correlation between $C_c$ and $\sir$) is explained by relative importance of spin-dependent scattering in solar WIMP capture and the fact that $\sigma_{SD,p}/\sigma_{SI,p}>10^3$ for most of the models in our set (cf. Figs. \ref{figs:sisd}-\ref{figs:eratio}).
    }
    \label{figs:elasticcap}
  \end{figure}

For models in equilibrium, the normalization of the raw flux spectra is determined by the capture rate, $\capr\,$. This relationship is illustrated in Figure \ref{figs:dcap}, where we display our models as points in the detected neutrino flux vs.\ capture rate plane. As expected, we see that the detected signal fluxes are well correlated with the capture rates for models in equilibrium (the remaining scatter of the points reflects the variance in the shape of the signal spectra), while the correlation is not as strong for out-of-equilibrium models.  The capture rate is, in turn, determined by a combination of $\sigma_{SI,p}$, $\sigma_{SD,p}$, $\Olsp$ and $m_{\LSP}$. Figure \ref{figs:elasticcap} illustrates the correlation between the capture rate and elastic scattering cross-sections. 

\begin{figure}[hbtp]
    \centering
    \includegraphics[width=0.8\textwidth]{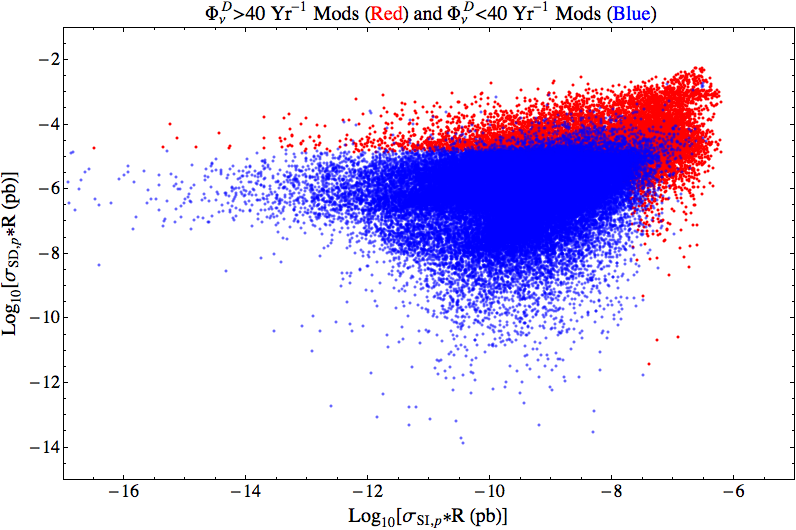}
    \caption{SUSY model dependence in the $\sdr$ vs.\ $\sir$ plane. Flat-prior models that are estimated to be excluded in the IC/DC search are displayed as red points, those that are not expected to be excluded are displayed as blue points.}
    \label{figs:sisd}
  \end{figure}

\begin{figure}[hbtp]
    \centering
    \includegraphics[width=1.0\textwidth]{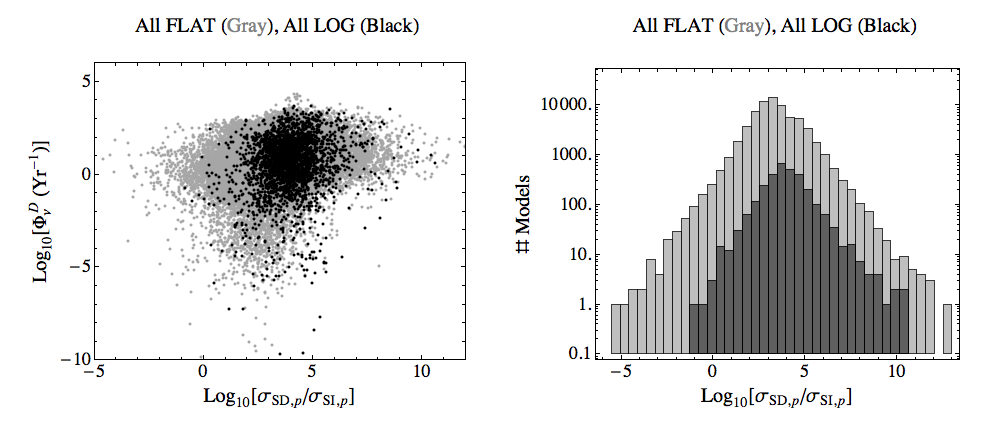}
    \caption{We describe the distribution of spin-dependent and spin-independent WIMP-proton cross-sections in our flat- and log-prior model sets. We display points in the detected neutrino flux, $\Phi^D_\nu$, vs.\ $\sigma_{SD,p}/\sigma_{SI,p}$ plane. We also histogram the ratio $\sigma_{SD,p}/\sigma_{SI,p}$ for both model sets. Flat-prior models are described by grey points/bars while log-prior models are described by black points/bars. We observe that most of the models in both sets have $\sigma_{SD,p}/\sigma_{SI,p}>10^3$ (the small subset of models for which $\sigma_{SD,p}<\sigma_{SI,p}$ are seen to have contributions to $\sigma_{SD,p}$ from $Z^0$ and squark diagrams which largely cancel).}
    \label{figs:eratio}
  \end{figure}

In Figure \ref{figs:elasticcap} it is apparent that the correlation between $\capr$ and $\sigma_{SD,p}$ is stronger than that between $\capr$ and $\sigma_{SI,p}$. This can be understood from Figures \ref{figs:sisd}-\ref{figs:eratio} as follows. In Figure \ref{figs:sisd} we observe the performance of the IC/DC search in the $\sdr$ vs.\ $\sir$ plane. We display models that would be expected to be excluded by the IC/DC analysis (red points) and also those that we would not expect to be excluded (blue points) in this plane and see that points that can be ruled out locate along approximate, but discernible, vertical and horizontal boundary lines on the figure. These lines roughly describe a reach of $\sdr\sim10^{-5}\; \rm{pb}$ and $\sir\sim10^{-8}\; \rm{pb}$ in this plane, and the ratio of these numbers implies that the IC/DC search is sensitive to spin-independent elastic scattering cross-sections that are roughly $\sim10^3$ times smaller than the thermal spin-dependent elastic scattering cross-sections to which it is sensitive (of course, terrestrial direct detection experiments currently probing $\sir$ have already surpassed this level of sensitivity). This ratio is essentially the ratio of the coefficients in Equation (\ref{captureeq}), which are, roughly, integrals over the solar volume of the abundances of various target nuclei\footnote{The precise size of this ratio is, of course, dependent on the specific choice of solar composition model. Uncertainties due to the solar model were discussed in \cite{Ellis:2009ka} and, while important in making accurate predictions, are not large enough to change the intuition of this discussion.}. The question becomes: what are typical values of the ratio $\sigma_{SD,p}/\sigma_{SI,p}$ in our pMSSM model set? The distribution of values for this ratio, in both our flat- and log- prior model sets, is displayed in Figure \ref{figs:eratio}. We find that $\sim66\%$ of flat-prior and $\sim86\%$ of log-prior models have $\sigma_{SD,p}/\sigma_{SI,p}>10^3$, so that for most of our models spin-dependent LSP-hydrogen scattering is the dominant mode of capture in the sun.

The model dependence that remains in Figures \ref{figs:dcap}-\ref{figs:elasticcap}, and the blurring of the horizontal and vertical ``lines" in Figure \ref{figs:sisd}, is due to the \emph{shape} of the signal neutrino spectrum, and its effect on the conversion of raw flux spectra to detected flux spectra. This conversion is performed as discussed above, by convolving raw fluxes with detector effective areas that are sharply energy dependent (decreasing with decreasing energy). The shape of the spectra is obviously highly dependent on pMSSM model details, most directly via the DM annihilation rates into various SM final states. 

\begin{figure}[hbtp]
    \centering
    \includegraphics[width=1.0\textwidth]{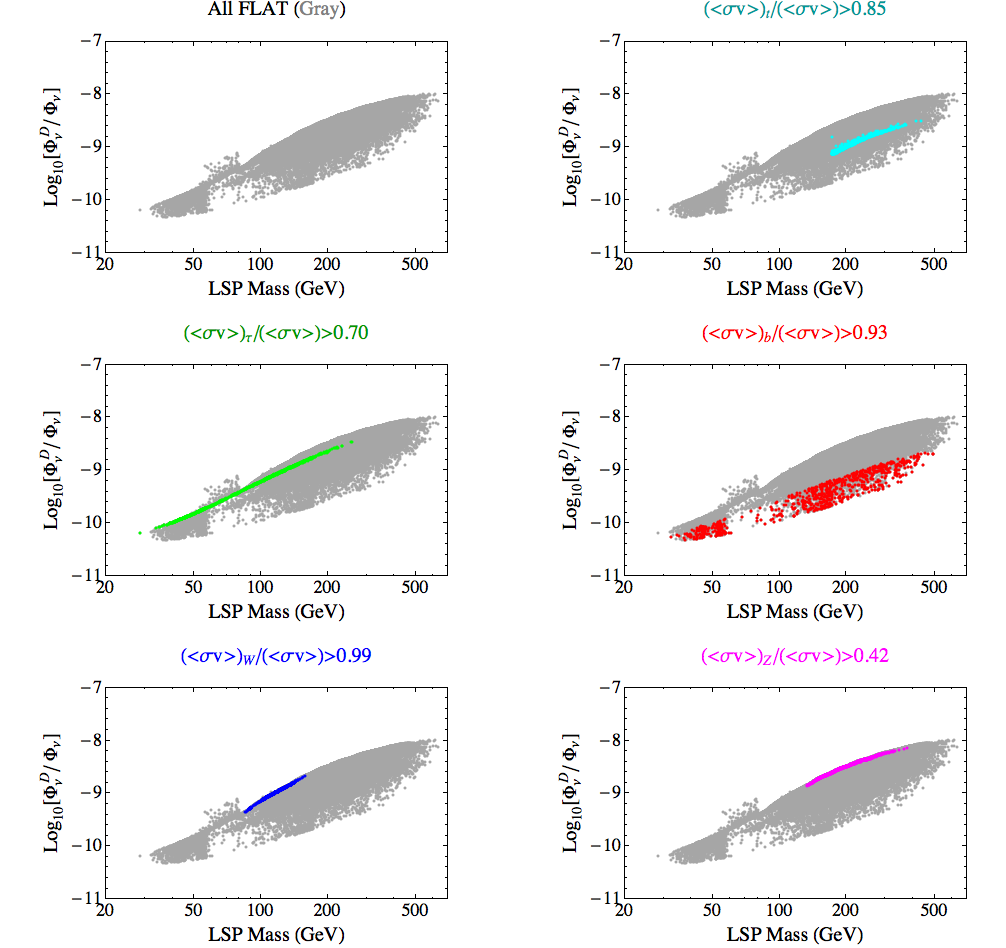}
    \caption{We display points representing our flat prior models in the $(\mathrm{Detected Flux})/(\mathrm{Raw Flux})$ (Eqn.~(\ref{convratio})) vs.\ LSP mass plane. The full flat-prior model set is displayed as grey points and models whose annihilations occur predominantly through a given final state channel are overlaid in other colors. Models with $\langle\sigma v\rangle_{t\bar{t}}/\langle\sigma v\rangle>0.85$ (cyan), with $\langle\sigma v\rangle_{\tau\bar{\tau}}/\langle\sigma v\rangle>0.70$ (green), with $\langle\sigma v\rangle_{b\bar{b}}/\langle\sigma v\rangle>0.93$ (red), with $\langle\sigma v\rangle_{W^+W^-}/\langle\sigma v\rangle>0.99$ (blue) and with $\langle\sigma v\rangle_{Z^0Z^0}/\langle\sigma v\rangle>0.42$ (magenta) are shown. Purities are chosen to obtain subsets of models of similar size (in our model set there is a maximum purity for annihilations into the $Z^0Z^0$ final state, $\langle\sigma v\rangle_{Z^0Z^0}/\langle\sigma v\rangle\leq0.445$, as the higgsino-like LSPs which annihilate well to $Z^0Z^0$ via $\tilde{\chi}^0_2$ exchange also annihilate well to $W^+W^-$ via $\chip$ exchange).}
    \label{figs:aeff}
  \end{figure}

\begin{figure}[hbtp]
    \centering
    \includegraphics[width=1.0\textwidth]{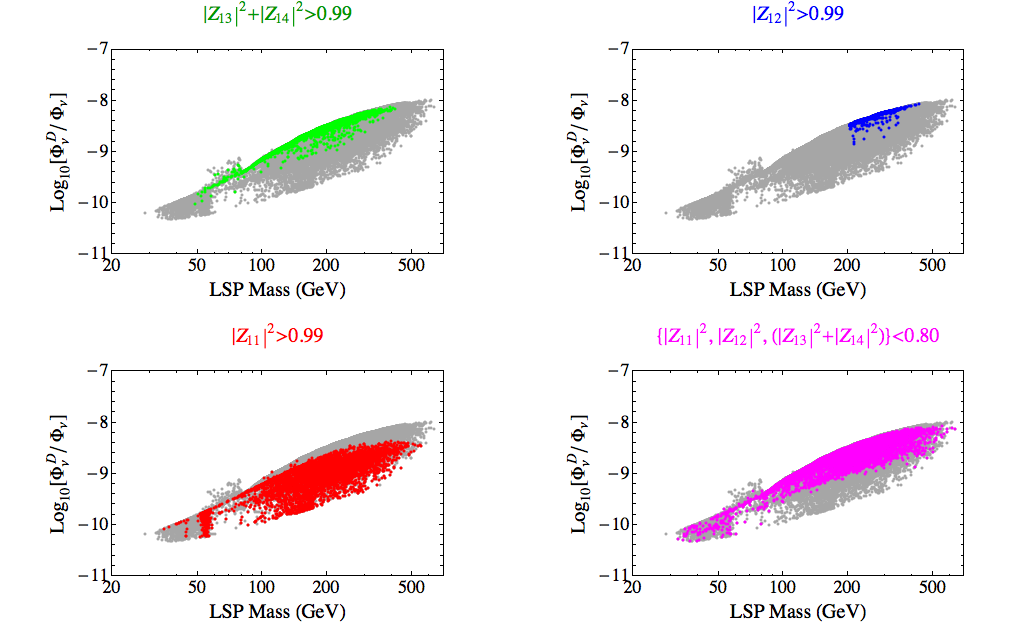}
    \caption{We display points representing our flat prior models in the $(\mathrm{Detected Flux})/(\mathrm{Raw Flux})$ (Eqn.~(\ref{convratio})) vs.\ LSP mass plane. The full flat-prior model set is displayed as grey points and models categorized according to LSP eigenstate composition are overlaid in other colors. By convention our LSPs are described in terms of their neutralino mass matrix entries as: $\LSP=Z_{11}\tilde{B}+Z_{12}\tilde{W}^3+Z_{13}\tilde{H}_1^0+Z_{14}\tilde{H}_2^0$. Higgsino models are defined as having $(|Z_{13}|^2+|Z_{14}|^2)>0.99$ and are displayed here in green. Wino models are defined as having $|Z_{12}|^2>0.99$ and are displayed here in blue. Bino models are defined as having $|Z_{11}|^2>0.99$ and are displayed here in red. Mixed models are defined as having $|Z_{11}|^2$, $|Z_{12}|^2$ and $(|Z_{13}|^2+|Z_{14}|^2)$ all $<0.8$ and are displayed here in magenta.}
    \label{figs:aeffino}
  \end{figure}

In order to study this dependence we focus on the ratio:
\begin{equation}
 \frac{(\mathrm{Detected Flux})}{(\mathrm{Raw Flux})} = \frac{\Phi^D_\nu}{(\Phi_{\nu_{\mu}}+\Phi_{\nu_{\bar{\mu}}})}=\frac{\int A^{\nu_\mu}_{\rm{eff}}(E)\frac{d\Phi_{\nu_\mu}}{dE} + A^{\overline{\nu}_\mu}_{\rm{eff}}(E)\frac{d\Phi_{\overline{\nu}_\mu}}{dE} dE}{(\Phi_{\nu_{\mu}}+\Phi_{\nu_{\bar{\mu}}})}.
     \label{convratio}
\end{equation}
In Figure \ref{figs:aeff} we display points in the $(\mathrm{Detected Flux})/(\mathrm{Raw Flux})$ vs.\ LSP mass plane for each pMSSM model in our flat-prior model set (grey), and for models that annihilate with various levels of purity (described in the caption) into particular SM final states. This ratio varies by about two orders of magnitude over this model set, generally increasing with increasing LSP mass, as the resulting neutrino spectra are shifted in energy toward larger effective areas. There is about an order of magnitude spread in the ratio of $(\mathrm{Detected Flux})/(\mathrm{Raw Flux})$ at any given LSP mass that is due to varying annihilation final states. As expected, the upper ``edge" of the scatter is made up of pMSSM models annihilating dominantly into so-called ``hard" channels, which yield relatively stiff neutrino spectra, $\ie$ $W^{+}W^{-}$, $Z^{0}Z^{0}$ and $\tau$-pairs. The lower edge is populated by models annihilating dominantly to $b\bar{b}$ and thus resulting in relatively soft neutrino spectra\footnote{The models with flux ratios that exceed the upper edge in the window $m_{\LSP}\sim 55-85 \gev$ are examples of LSPs whose annihilations are largely to the $\gamma Z^0$ final state, resulting in a neutrino spectrum from the $Z^0$ decay that is even stiffer than the spectra that would result from annihilation into the $Z^{0}Z^{0}$ final state (these models nevertheless end up with small signal rates because of generally low relic density).}. The correlation between LSP eigenstate composition and this ratio, $(\mathrm{Detected Flux})/(\mathrm{Raw Flux})$, is not as strong as that between annihilation final states and this ratio (cf. Figs. \ref{figs:aeff} and \ref{figs:aeffino}) as the connection between the two is muddied somewhat by the details of the superpartner mass spectrum.  Despite this, some amount of correlation is still evident. Nearly pure higgsino and wino LSPs typically annihilate with large rates into the $W^{+}W^{-}$ and $Z^{0}Z^{0}$ final states, thus tending to populate the upper edge in the figures. Nearly pure bino LSPs annihilate dominantly through helicity suppressed sfermion exchange graphs so that (for $m_{\LSP}\leq m_{t}$) annihilations of bino LSPs proceed to a mixture of $b\bar{b}$ and $\tau$-pair final states, and the resulting $(\mathrm{Detected Flux})/(\mathrm{Raw Flux})$ ratios interpolate between the pure $\tau$ and pure $b$ results. 

\begin{figure}[hbtp]
    \centering
    \includegraphics[width=1.0\textwidth]{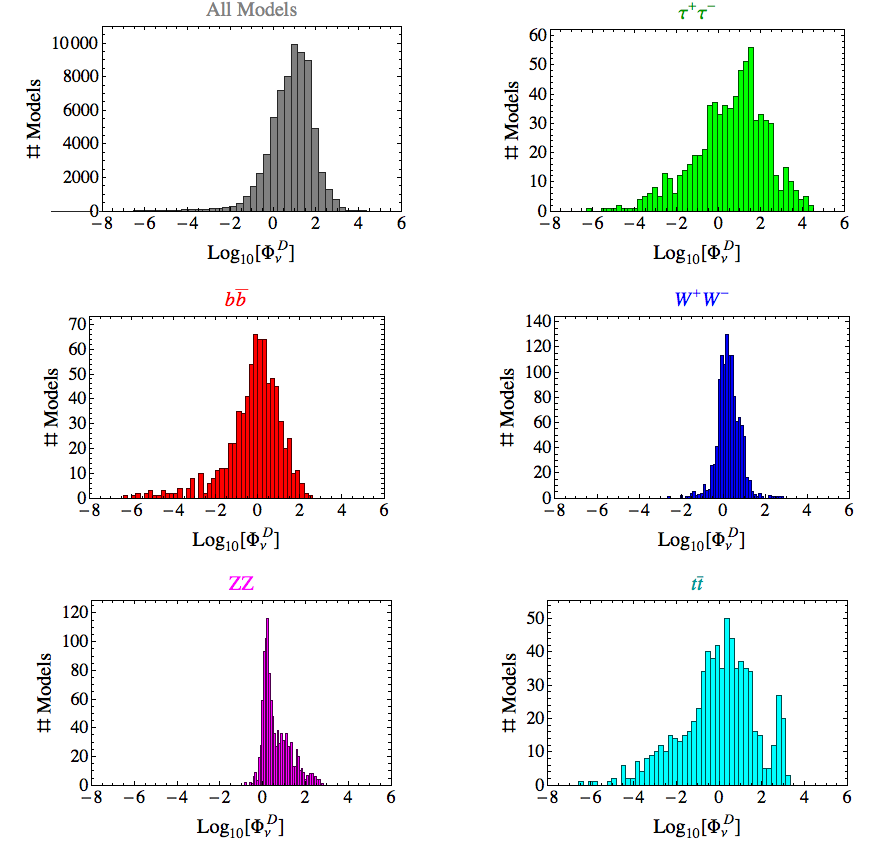}
    \caption{Detected neutrino flux histograms illustrating the dependence of $\Phi^D_{\nu}$ on final state annihilation channel. The full flat-prior model set is described in the grey histogram while subsets of these models whose annihilations occur predominantly through a given final state channel are displayed in other colors. Models with $\langle\sigma v\rangle_{t\bar{t}}/\langle\sigma v\rangle>0.85$ (cyan), with $\langle\sigma v\rangle_{\tau\bar{\tau}}/\langle\sigma v\rangle>0.70$ (green), with $\langle\sigma v\rangle_{b\bar{b}}/\langle\sigma v\rangle>0.93$ (red), with $\langle\sigma v\rangle_{W^+W^-}/\langle\sigma v\rangle>0.99$ (blue) and with $\langle\sigma v\rangle_{Z^0Z^0}/\langle\sigma v\rangle>0.42$ (magenta) are shown. Purities are chosen to obtain subsets of models of similar size (in our model set there is a maximum purity for annihilations into the $Z^0Z^0$ final state, $\langle\sigma v\rangle_{Z^0Z^0}/\langle\sigma v\rangle\leq0.445$, as the higgsino-like LSPs which annihilate well to $Z^0Z^0$ via $\tilde{\chi}^0_2$ exchange also annihilate well to $W^+W^-$ via $\chip$ exchange).}
    \label{figs:histofs}
  \end{figure}

\begin{figure}[hbtp]
    \centering
    \includegraphics[width=1.0\textwidth]{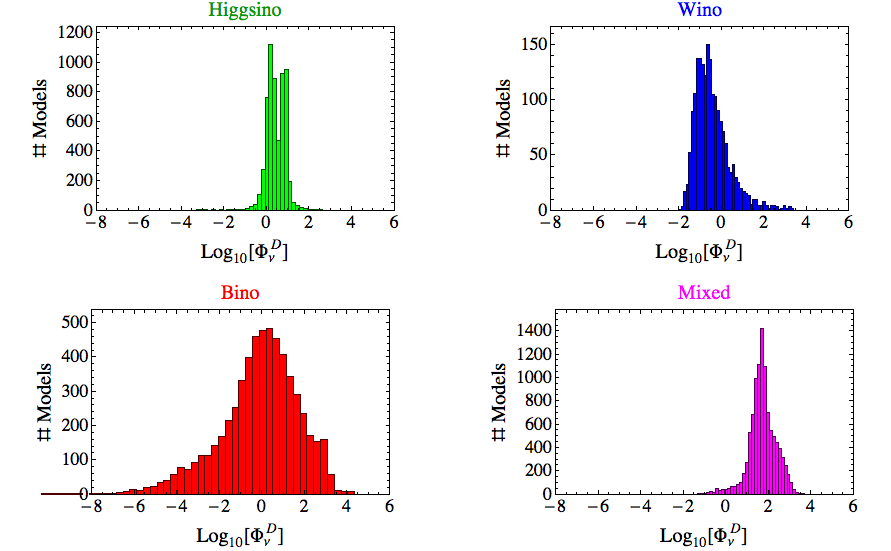}
    \caption{ Detected neutrino flux histograms illustrating the dependence of $\Phi^D_{\nu}$ on LSP eigenstate composition. By convention our LSPs are described in terms of their neutralino mass matrix entries as: $\LSP=Z_{11}\tilde{B}+Z_{12}\tilde{W}^3+Z_{13}\tilde{H}_1^0+Z_{14}\tilde{H}_2^0$. Higgsino models are defined as having $(|Z_{13}|^2+|Z_{14}|^2)>0.99$ and are displayed here in green. Wino models are defined as having $|Z_{12}|^2>0.99$ and are displayed here in blue. Bino models are defined as having $|Z_{11}|^2>0.99$ and are displayed here in red. Mixed models are defined as having $|Z_{11}|^2$, $|Z_{12}|^2$ and $(|Z_{13}|^2+|Z_{14}|^2)$ all $<0.8$ and are displayed here in magenta.}
    \label{figs:histoino}
  \end{figure}

In Figures \ref{figs:histofs}-\ref{figs:histoino} we show the detected flux histograms for various annihilation final states and LSP eigenstate compositions. One particularly interesting result is that the detected fluxes for mixed neutralinos are found to be typically quite high. This can be seen as the consequence of several factors. The most important factor in attaining high fluxes is the combination $\sdr$. The spin-dependent cross-section itself is due to $Z^{0}$ or squark mediated graphs. In any LSP eigenstate scenario the squark mediated contributions to scattering exist and are largely dependent on the scanned squark masses. However, potentially\footnote{In principle the $Z^{0}$ and squark contributions, with associated signed nuclear form factors, may add constructively or cancel against each other, depending on the details of the model spectrum and assumed values for the nuclear form factors.} larger cross-sections exist in cases where the LSP has significant higgsino fraction, and thus can couple strongly to the $Z^{0}$. As a competing effect, a large higgsino content opens several new annihilation channels and has been shown in simplified models to give LSP relic densities that fall significantly short of the WMAP measured value (for LSPs below $\sim1\tev$ \cite{ArkaniHamed:2006mb}). The extent to which lower relic density implies a lower overall value of $\sdr$ depends upon how closely the present day elastic scattering cross-sections are related to the freeze-out annihilation cross section. Annihilation sub-processes such as $\LSP\LSP\to h,H, Z^0 \to q\bar{q}$ decrease the relic density while simultaneously (potentially) increasing SI and SD scattering rates. In contrast, annihilations to non-quark final states only deplete the LSP relic density (resonant funnels are also far more efficient in s-channel annihilations than in t-channel elastic scattering). Finally, as noted above, higgsino-like LSPs may have significant annihilation rates into the $W^{+}W^{-}$ and $Z^{0}Z^{0}$ final state channels, $\ie$, those which see the largest detector effective area. LSPs with sizeable wino content share this latter virtue (annihilation to hard channels), but because of the lack of couplings to $h,Z^{0}$, as well as small $\chip-\LSP$ mass splittings which lead to efficient annihilation, they do not usually have high values of $\sdr$. A look into the subset of mixed LSPs in our flat-prior model set shows that essentially all models have a significant higgsino fraction ($97\%$ of the mixed models have LSPs that are $>20\%$ higgsino), so that many of these models yield observable solar neutrino rates. 
  
  \subsection{IC/DC Comparison with Other Experiments}
  \label{sec:excomps}
  
  \indent We now compare the ability of IC/DC to constrain or discover the pMSSM models in our set with that of direct detection experiments, searches in other indirect detection experiments, and at the $7\tev$ LHC. In order to make this comparison, we focus on constraints in the $\sir$ vs.\ LSP mass, $\sdr$ vs.\ LSP mass and $\sir$ vs.\ $\sdr$ planes. In the following figures, pMSSM models which are expected to be excluded by the IC/DC 1825 day dataset ($\ie$, those with detected neutrino fluxes $\Phi^D_{\nu}>40\:\rm{yr}^{-1}$) are displayed as red points over the grey points which represent the entire flat-prior model set. pMSSM models which cannot be excluded are displayed in blue.
  
\begin{figure}[hbtp]
    \centering
    \includegraphics[width=1.0\textwidth]{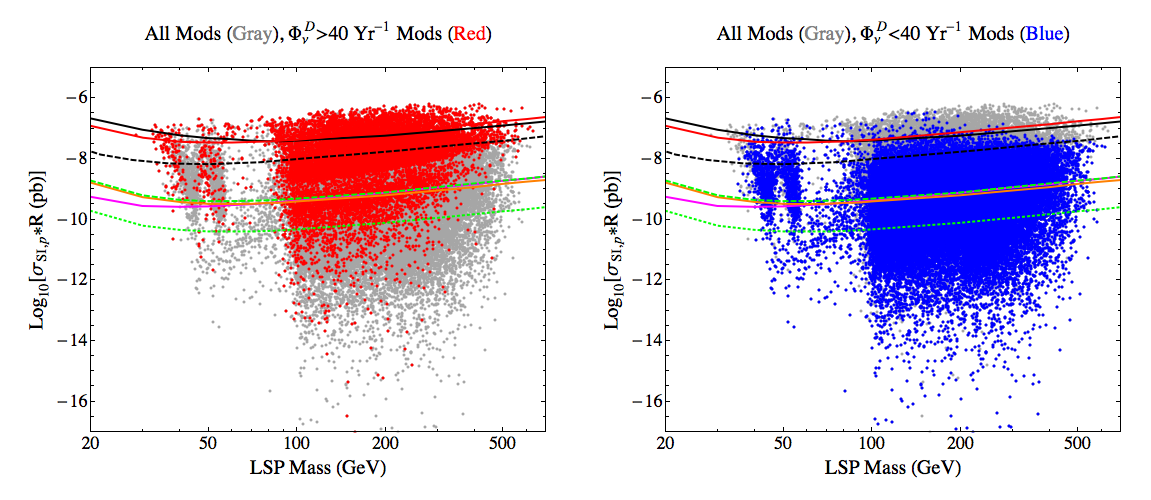}
    \caption{Comparison of IC/DC and spin-independent direct detection searches. We display all points in the flat-prior model set in grey, models that are estimated to be excluded by the IC/DC solar WIMP search in red and those which are not excluded in blue. Current experimental limits from the CDMS \cite{Ahmed:2009zw} and XENON100 \cite{Aprile:2010um}\cite{Aprile:2011hi}collaborations are displayed as red (CDMS 2010), black-solid (XENON100 2010) and black-dashed (XENON100 2011), respectively. Near-future projected experimental limits from LUX \cite{McKinsey:2010zz}, SuperCDMS \cite{Bruch:2010eq}, COUPP 60kg and COUPP 500kg \cite{bigcoupp} are displayed as brown, magenta, orange, green-dashed and green-dotted lines, respectively.}
    \label{figs:siexperiments}
  \end{figure}

\begin{figure}[hbtp]
    \centering
    \includegraphics[width=1.0\textwidth]{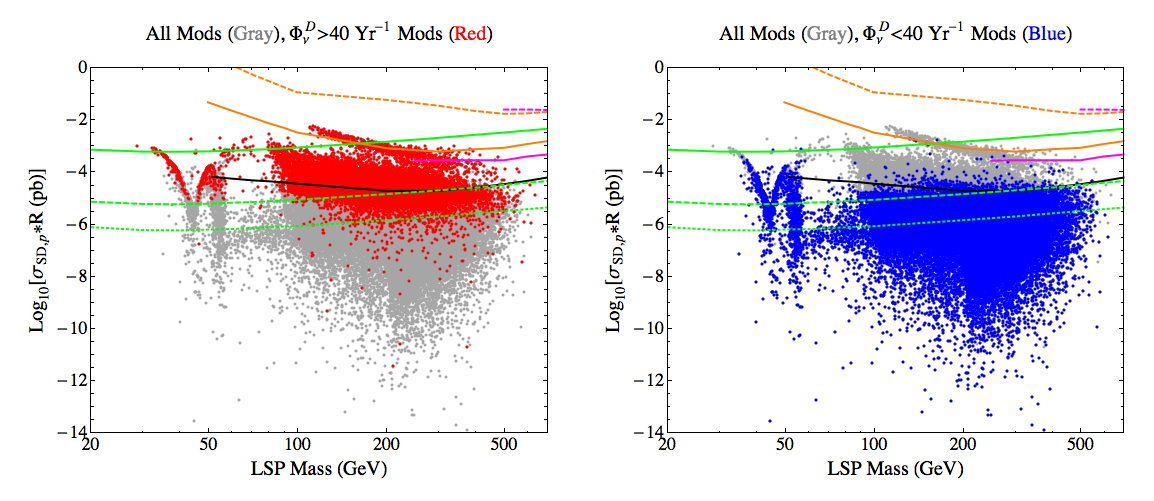}
    \caption{
    Comparison of IC/DC and spin-dependent direct detection searches. We display all points in the flat-prior model set in grey, models that are estimated to be excluded by the IC/DC solar WIMP search in red and those which are not estimated to be excluded in blue. Current experimental limits from the AMANDA \cite{Braun:2009fr} and IceCube-22 \cite{Abbasi:2009uz} collaborations are displayed as orange and magenta lines, respectively (with the assumption of soft or hard channel annihilations represented by dashed or solid lines, respectively). Near-future projected experimental limits from the COUPP \cite{Behnke:2010xt}\cite{bigcoupp} 4kg, 60kg and 500kg searches in green- solid, dashed and dotted lines, respectively. The IceCube/DeepCore limit estimated in \cite{Wiebusch:2009jf} (assuming hard channel annihilations, DM which is in capture/annihilation equilibrium and DM which has $R=1$) is displayed as a black line. 
     }
    \label{figs:sdexperiments}
  \end{figure}

  For comparison to current and planned direct detection experiments we combine the neutrino fluxes calculated in this work with quoted limits on the thermal spin-independent cross-section, $\sir$, from CDMS \cite{Ahmed:2009zw} and XENON100 \cite{Aprile:2010um}, as well as the recent XENON100 limit \cite{Aprile:2011hi}. We also show projected limits on $\sir$ from LUX (30,000 kg/days) \cite{McKinsey:2010zz}, SuperCDMS at SNOLAB \cite{Bruch:2010eq} and COUPP (60kg and 500kg 1yr searches \cite{bigcoupp}) in Figure \ref{figs:siexperiments}. Here, we recall that in generating our pMSSM model set \cite{Berger:2008cq}, we employed the direct detection limits that were applicable at the time as a constraint on viable models (while allowing a factor of 4 error to account for the uncertainties associated with computing the spin-independent LSP-nucleonic scattering cross-section). The recent XENON100 result \cite{Aprile:2011hi} represents the current most stringent limit on $\sir$ and certainly rules out some subset of our pMSSM model set. While we have not dropped these models from the present discussion, we will consider the impact of this result in what follows. For the thermal spin-dependent scattering cross-section, $\sdr$, we show the quoted limits from AMANDA (2001-2003 data) \cite{Braun:2009fr} and the IceCube-22 string (no DeepCore) 2007 data \cite{Abbasi:2009uz}, we also display projected limits on $\sdr$ from COUPP (4kg \cite{Behnke:2010xt}, 60kg and 500kg \cite{bigcoupp} 1yr searches) and IceCube/DeepCore \cite{Wiebusch:2009jf} in Figure \ref{figs:sdexperiments}. Recall that limits from Amanda and IceCube are placed under the assumption of annihilation dominantly to either hard or soft final state channels. 
  
  While there is substantial overlap between excluded and non-excluded models in both of the quantities $\sir$ and $\sdr$, we observe that essentially all of pMSSM models with $\sir>10^{-7}\; \rm{pb}$ or $\sdr>10^{-4}\; \rm{pb}$, and a majority of models with $\sir>10^{-8}\; \rm{pb}$ or $\sdr>10^{-5}\; \rm{pb}$, would be excluded by the IC/DC search. The recent XENON100 limit (the black-dashed curve in Fig.\ \ref{figs:siexperiments}) is $\sim10^{-8}\; \rm{pb}$ over most of the LSP masses in our set. This value excludes $\sim16\%$ of the models in our flat-prior set, although we note that $\sim10^{-8}\; \rm{pb}$ lies on the steep portion of the $\sir$ distribution for these models so that, if nuclear form factors were to conspire so that the appropriate limit were looser or tighter by a factor of 2, $\sim10\%$ or $\sim25\%$ of our flat-prior models would be excluded, respectively. If we assume, as a rough estimate, that all models with $\sir>10^{-9}\; \rm{pb}$ will be excluded in the near-future ton-scale spin-independent scattering direct searches, we estimate that, in our flat-prior model set, $\sim$18\% of the models will be excluded by both the IC/DC search and the spin-independent direct detection (SI DD) searches, while $\sim$4\% will be excluded by IC/DC and not by SI DD searches, $\sim$31\% will be excluded by SI DD searches but not by IC/DC and $\sim$47\% of these models would not be excluded by either IC/DC or SI DD searches. 
  
Recall that the previously estimated IC/DC exclusion limit (the black line in Figure \ref{figs:sdexperiments}, based on the derivation presented in in \cite{Wikstrom:2009kw}) assumes \emph{(i)} annihilations exclusively to hard channels, \emph{(ii)} an LSP in capture/annihilation equilibrium and \emph{(iii)} an LSP that makes up all of DM ($R=1$). One can see that this line agrees fairly well with the putative upper line described by the set of our models which IC/DC cannot exclude (the blue points in Figure \ref{figs:sdexperiments}); only a small number of such models have $\sdr$ above this line. Upon examination these are models which annihilate dominantly to soft channels (violating assumption \emph{(i)} above) and, for non-excluded models well above the line, a significant amount of squark coannihilation is present so that the scattering cross sections are relatively large (due to the light squark) but still difficult to constrain as $R<1$ (violating assumption \emph{(ii)} above). As expected there are a large number of models that may be excluded with $\sdr$ ranging much below this line: these are models for which $C_c$ is dominated by spin-independent scattering. A large fraction of these models would thus also be expected to be excluded by SI DD experiments.

  We note that the IC/DC and COUPP 60kg searches seem to do a similar job of probing \emph{both} $\sdr$ \emph{and} $\sir$, although it may not be obvious from the information provided in Figs.\ \ref{figs:siexperiments}-\ref{figs:sdexperiments} alone. From Figure \ref{figs:sdexperiments} there are apparently quite a few models with $\sdr$ far below the sensitivity expected by COUPP that may indeed be excluded by IC/DC. However, these are necessarily models for which $C_c$ receives a sizable contribution from $\sir$ as discussed above. We would expect that $\sir\gsim10^{-9}$ pb for such models, roughly within the $\sir$ sensitivity expected of the COUPP 60kg search. We thus expect the two experiments to provide highly complementary cross-checks on any WIMP discovery or limits in the $\sdr$ vs.\ $\sir$ plane.
  
  For comparison to potential discoveries at the LHC we employ the expected results of an ATLAS 4j0l ($\ie$, 4-jet + 0-lepton + missing transverse energy) search ($\sqrt{s}=7\tev$, $1\:\mathrm{fb}^{-1}$ and $50\%$ systematic uncertainty are assumed) that were found for this same set of pMSSM models in the work \cite{Conley:2011nn}\footnote{Results for other search channels are similar. We use the 4j0l search channel here as it was found to be the most effective discovery channel for the pMSSM \cite{Conley:2011nn}.}. Given that models characterized as ``passing" or ``failing" the ATLAS analysis was based on a discovery criterion, rather than an exclusion criterion, we compare these results to our (conservative) criterion for discovery in IC/DC, $\Phi^D_{\nu}=100~\rm{events/yr}$ (see Section \ref{sec:icdcsearch}). Using this criterion, we note that $\sim8.6\%$ of the models in our flat-prior pMSSM set are expected to be discoverable in the IC/DC solar WIMP search.
    
\begin{figure}[hbtp]
    \centering
    \includegraphics[width=1.0\textwidth]{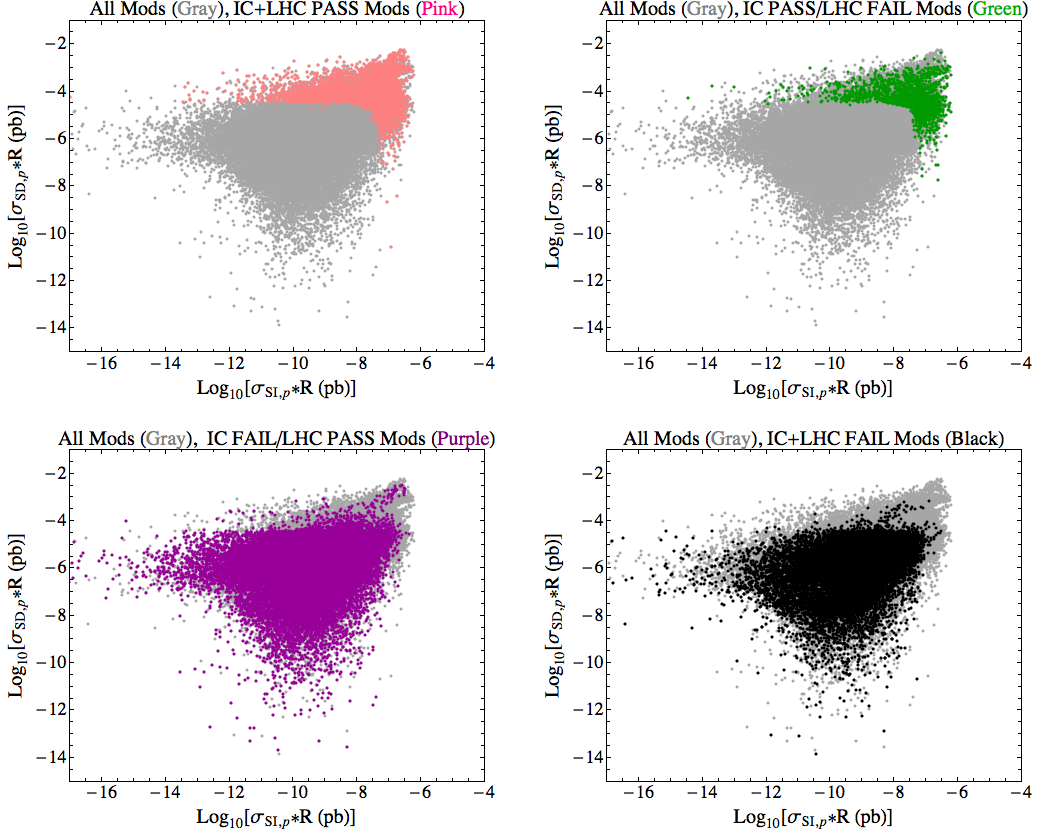}
    \caption{
    Comparison of IC/DC and the ATLAS 4j0l ($\sqrt{s}=7\tev$, $1\:\mathrm{fb}^{-1}$ and $50\%$ systematic uncertainty assumed) searches. Here ``PASS" and ``FAIL" denote discovered or not discovered, respectively. We display all points in the flat-prior model set in grey, models that are estimated to be discoverable by both the IC/DC solar WIMP search and the ATLAS search are displayed in pink, those expected to be seen by IC/DC but missed in the ATLAS search in green, those expected to be missed by IC/DC but seen in the ATLAS search in purple and those which are estimated to be unobservable by either search are displayed in black.}
    \label{figs:sisdlhc}
  \end{figure}

  Figure \ref{figs:sisdlhc} demonstrates the combined ability of the IC/DC solar WIMP search and the ATLAS 4j0l search to discover models in our flat-prior pMSSM model set. In this figure we display models that would be expected to be seen by both searches, as well as models that would be discovered at IC/DC and missed in the 4j0l search, models that would be missed at IC/DC and seen in the 4j0l search and finally models expected to be missed by both searches in the $\sdr$ vs.\ $\sir$ plane. Of course, in the time that will be required for IC/DC to perform the search described in this paper, the LHC will likely accumulate much more than $1\:\mathrm{fb}^{-1}$ of data. However, the expected coverage of our model set by LHC searches using , $\eg$, $10\:\mathrm{fb}^{-1}$ of data, is $>90\%$ \cite{Conley:2011nn} (the vast majority of our pMSSM models would be seen at the LHC) and it makes little sense to ask which models are seen or unseen by combinations of IC/DC and the LHC in this scenario. The point of Figure \ref{figs:sisdlhc} is not to compare coverage of the two classes of experiment at some definite time in the future, but to observe that the shape of these regions are determined primarily by the IC/DC search (cf. Fig. \ref{figs:sisd}). This is illustrative of the fact that the solar WIMP signal is very strongly correlated with the elastic scattering cross-sections while the rate of 4j0l events in ATLAS is, of course, only very indirectly so. 
  
  
  As a final remark, we discuss the relative performance of searches sensitive to the DM \emph{annihilation} cross-section, $\sigv$; a detailed study of such signals from DM in this pMSSM model set was investigated in {\cite{Cotta:2010ej}. As mentioned before, in contrast to searches in which signals are largely determined by elastic scattering cross sections ($\ie$, in direct detection experiments and the IC/DC solar WIMP search), indirect searches for DM annihilation will probe signals $\sim\!\!\tsigv$. In lieu of non-standard cosmological scenarios \cite{resDM}\cite{gordy}, which may drastically alter the connection between the annihilation cross-section and the relic density, or special relationships in the SUSY mass spectrum that allow for co-annihilations or resonant annihilations, we expect the scaling $R\sim1/\sigv$, so that $\tsigv\sim R$ while $\sdr\sim1$.  Thus indirect searches for DM annihilation that probe $\tsigv$, such as the PAMELA/FERMI/AMS-02 cosmic-ray antimatter measurements, FERMI/MAGIC/HESS $\gamma$-ray measurements and IC/DC observations of neutrinos from the Galactic Center, are much more sensitive to the LSP relic density than the other classes of experiments. Such experiments will typically have a much harder time discovering LSPs with $R\ll1$. We note that the largest annihilation cross sections in our flat-prior set are $\tsigv\sim 6\times 10^{-26}\:\rm{cm}^3\:\rm{s}^{-1}$, and only about a quarter of the models in this set have $\tsigv\geq10^{-27}\:\rm{cm}^3\:\rm{s}^{-1}$. While astrophysical indirect detection limits typically come with much larger uncertainties than other classes of experiments, due to the difficulty in estimating the strength of the annihilation source ($\eg$, in estimating DM halo profiles and substructure \cite{substructure}), a sensitivity at the level of $\tsigv\geq10^{-27}\:\rm{cm}^3\:\rm{s}^{-1}$ or better, for canonical choices of profile, etc., will probably be necessary in order to be sensitive to a large fraction of models in our set.
 
  \section{Discussion}
  \label{sec:discuss}
 
 In this paper we have investigated the ability of the upcoming 1825 day IceCube/DeepCore solar WIMP search to discover/constrain SUSY WIMPs. In this aim we have employed the large set of $\sim71$k phenomenologically-viable pMSSM SUSY models that were generated and described in the work \cite{Berger:2008cq}. We have discussed the basic calculation of neutrino telescope signals from captured WIMPs annihilating in the solar core and the details of our analysis, which relied heavily on the use of the computational package DarkSUSY \cite{Gondolo:2004sc}. 
 
 We have discussed sources of uncertainty that affect the capture process.  Many works have previously elucidated important sources of experimental uncertainty \cite{formfactors}\cite{Ellis:2009ka} ($\ie$, from our imprecise knowledge of nuclear matrix elements, solar composition and of the details of the neutrino sector) in the estimation of the capture rate and subsequent neutrino signal. Although we made no attempt to account for such uncertainties here, we have found that, given the diversity of our pMSSM model spectra, much larger errors (orders of magnitude) in the resulting rates would result from a poor choice of parton-level scattering amplitude calculations and/or failure to take into account the IC/DC detector effective area. Both of these have been carefully considered in this work. 
 
As a basic result of our analysis, we find that a large fraction of our pMSSM models are expected to have significant signal rates in this search. We find that LSPs with a wide range of masses can be excluded and, somewhat surprisingly, the IC/DC search reach extends to many of the models for which the LSP forms only a small fraction of the total DM abundance. A study of solar capture/annihilation equilibrium confirms the expected result, that essentially none of the out-of-equilibrium LSPs in our set are are expected to be excluded. We have compared the results from subsets of pMSSM models that were generated with flat- and log-prior scanning over parameters and have found them to be quite similar. 

We have described the SUSY model dependence that is seen in the determination of both the shape and normalization of the resulting signal neutrino spectra. Since most all of the models that may be excluded by IC/DC are in capture/annihilation equilibrium the normalization of the spectra is essentially determined by the capture rate and since most of our models have $\sigma_{SD,p}/\sigma_{SI,p}>10^3$ and the sun is largely composed of hydrogen nuclei targets, the normalization is mostly determined by $\sdr$. The shape of the spectra are of course most closely tied to the annihilation final state channels and we find that, for a given LSP mass, the final state channels producing the hardest and softest neutrino spectra see about an order of magnitude difference in the effective area of the detector.

Given the relative importance of the SD cross-section, relic density and final state annihilation channel we find semi-predictive differences between the results expected from classes of models with bino, wino, higgsino and mixed LSP eigenstates. We observe that nearly pure wino or higgsino LSP models typically have low (though not always, unobservable) rates due to their generally very low LSP relic densities. The bino LSP models have the widest range of predictions as both their relic density and elastic scattering cross sections are largely determined by the scanned sfermion masses. We noted that mixed LSP models typically predict very large rates and understand this as arising from their LSPs almost always having a significant higgsino fraction. Such models can thus attain large SD scattering cross sections and large annihilation rates into hard channel final states, without the very low relic densities that would result from being very purely higgsino.

We conclude with our expectation that the IceCube/DeepCore search may play an integral role in the experimental confirmation or expulsion of Supersymmetric neutralino dark matter. In the event of production and measurement of new invisible states at the LHC there can be no guarantee of their cosmological stability or relic abundance until complementary observations can be made in direct or indirect detection experiments. While terrestrial direct detection experiments are already poised to probe deep into the space of scattering cross sections populated by our models, the IC/DC search offers competitive sensitivity to the same scattering cross sections on a similar time scale and with somewhat orthogonal systematics.  
 
  \section{Acknowledgements}

  The authors would like to thank J. Conley, C. de los Heros, J. Edsj$\ddot{\rm{o}}$, J. Gainer, D. Grant, P. Gondolo, T. Montaruli and K. Olive for discussions related to this work. The work of R.C.C. is supported in part by an NSF Graduate Fellowship.


\end{document}